\documentclass[12pt]{article}
\usepackage{fullpage}
\usepackage{epsf}
\usepackage{epsfig}
\usepackage{amsthm}
\usepackage{amsfonts}
\usepackage{color}
\usepackage{amsmath}
\usepackage{setspace}
\usepackage{natbib}
\usepackage{hyperref}
\usepackage{algorithmic}
\usepackage[boxed]{algorithm}
\usepackage[usenames, dvipsnames]{xcolor}

\usepackage{graphicx}
\usepackage{caption}
\usepackage{subcaption}
\usepackage[normalem]{ulem}

\newcommand{\blind}{0}

\DeclareMathOperator{\Diag}{Diag}
\DeclareMathOperator{\tr}{tr}

\begin{document}

\newcommand{\bm}[1]{\mbox{\boldmath $#1$}}
\newcommand{\mb}[1]{#1}
\newcommand{\bE}[0]{\mathbb{E}}
\newcommand{\bV}[0]{\mathbb{V}\mathrm{ar}}
\newcommand{\bP}[0]{\mathbb{P}}
\newcommand{\ve}[0]{\varepsilon}
\newcommand{\mN}[0]{\mathcal{N}}
\newcommand{\iidsim}[0]{\stackrel{\mathrm{iid}}{\sim}}
\newcommand{\NA}[0]{{\tt NA}}
\newcommand{\cB}{\mathcal{B}}
\newcommand{\R}{\mathbb{R}}
\newcommand{\Rp}{\R_+}

\newcommand{\x}{\mathbf{x}}
\newcommand{\xu}{\bar{\mathbf{x}}} 
\newcommand{\yu}{\bar{y}} 
\newcommand{\z}{\mathbf{z}}
\newcommand{\veck}{\mathbf{k}}
\newcommand{\veckN}{\mathbf{k}_N}
\newcommand{\veckn}{\mathbf{k}_n}
\newcommand{\veccn}{\mathbf{c}_n}
\newcommand{\veccN}{\mathbf{c}_N}
\newcommand{\vecX}{\mathbf{X}}
\newcommand{\vecXu}{\bar{\mathbf{X}}} 
\newcommand{\vecY}{\mathbf{Y}}
\newcommand{\vecYu}{\bar{\mathbf{Y}}} 
\newcommand{\A}{\mathbf{A}}
\newcommand{\An}{\mathbf{A}_n}
\newcommand{\B}{\mathbf{B}}
\newcommand{\C}{\mathbf{C}}
\newcommand{\Cg}{\mathbf{C}_{(g)}}
\newcommand{\CN}{\mathbf{C}_N}
\newcommand{\Cn}{\mathbf{C}_n}
\newcommand{\D}{\mathbf{D}}
\newcommand{\Dn}{\mathbf{D}_n}
\newcommand{\DN}{\mathbf{D}_N}

\newcommand{\Deltan}{\boldsymbol{\Delta}_n}
\newcommand{\I}{\mathbf{I}}
\newcommand{\K}{\mathbf{K}}
\newcommand{\KN}{\mathbf{K}_N}
\newcommand{\Kn}{\mathbf{K}_n}
\newcommand{\SN}{\boldsymbol{\Sigma}_N}
\newcommand{\Sn}{\boldsymbol{\Sigma}_n}

\newcommand{\nug}{\nu_{(g)}}
\newcommand{\hnug}{\hat{\nu}_{(g)}}

\newcommand{\Un}{\boldsymbol{\Upsilon}_n}
\newcommand{\Ug}{\boldsymbol{\Upsilon}_{(g)}}

\newcommand{\U}{\mathbf{U}}
\newcommand{\V}{\mathbf{V}}
\newcommand{\ones}{\bm{1}}
\newcommand{\Lan}{\boldsymbol{\Lambda}_n}
\newcommand{\LaN}{\boldsymbol{\Lambda}_N}
\newcommand{\nun}{\hat{\nu}_n}
\newcommand{\nuN}{\hat{\nu}_N}
\newcommand{\Or}{\mathcal{O}}

\newtheorem{lemma}{Lemma}[section]
\newtheorem{Proposition}{Proposition}[section]
\theoremstyle{remark}
\newtheorem{remark}{Remark}[section]

\newcommand{\blu}[1]{\textcolor{black}{#1}}


\if0\blind
{
\title{\vspace{-1.25cm} Practical heteroskedastic Gaussian process modeling for large simulation experiments}
\author{Micka\"{e}l Binois\thanks{Corresponding author: The University of
Chicago Booth School of Business, 5807 S.~Woodlawn Ave., Chicago IL, 60637;
\href{mailto:mbinois@chicagobooth.edu}{\tt mbinois@chicagobooth.edu}}
\and
Robert B.~Gramacy\thanks{Department of Statistics, Virginia Tech, Hutcheson Hall,
250 Drillfield Drive, Blacksburg, VA 24061}
\and Mike Ludkovski\thanks{Department of Statistics and Applied Probability,
University of California Santa Barbara, 5520 South Hall Santa Barbara, CA 93106-3110}
}

\date{}

\maketitle
}\fi

\if1\blind
{
  \bigskip
  \bigskip
  \bigskip
  \begin{center}
    {\LARGE\bf Practical heteroskedastic Gaussian process modeling for large simulation experiments}
\end{center}
  \medskip
  \bigskip
} \fi

\begin{abstract}
We present a unified view of likelihood based Gaussian progress regression for
simulation experiments exhibiting input-dependent noise. Replication plays an
important role in that context, however previous methods leveraging replicates
have either ignored the computational savings that come from such design, or
have short-cut full likelihood-based inference to remain tractable.  Starting
with homoskedastic processes, we show how multiple applications of a
well-known Woodbury identity facilitate inference for all parameters under
the likelihood (without approximation), bypassing the typical full-data sized
calculations.  We then borrow a latent-variable idea from machine learning to
address heteroskedasticity, adapting it to work within the same thrifty
inferential framework, thereby simultaneously leveraging the computational and
statistical efficiency of designs with replication.  The result is an
inferential scheme that can be characterized as single objective function,
complete with closed form derivatives, for rapid library-based optimization.
Illustrations are provided, including real-world simulation experiments from
manufacturing and the management of epidemics.

\end{abstract}


\doublespacing

\section{Introduction}
\label{sec:intro}

Simulation-based experimentation has been commonplace in the physical and
engineering sciences for decades, and has been advancing rapidly in the social
and biological sciences.  In both settings, supercomputer resources
have dramatically expanded the size of experiments.  In the physical/engineering
setting, larger experiments are desired to enhance
precision and to explore larger parameter spaces.  In the social and
biological sciences there is a third reason:  stochasticity.  Whereas in the
physical sciences solvers are often deterministic, or if they involve Monte
Carlo then the rate of convergence is often known
\citep{picheny:ginsbourger:2013}, in the social and biological sciences
simulations tend to involve randomly interacting agents.
In that setting, signal-to-noise ratios can
vary dramatically across experiments and for configuration (or input) spaces
within experiments.  We are motivated by two examples, from inventory
control \citep{hong:nelson:2006,xie:frazier:chick:2012} and online management of
emerging epidemics \citep{hu2015sequential}, which exhibit both features.

Modeling methodology for large simulation efforts with intrinsic stochasticity
is lagging. One attractive design tool is \emph{replication}, i.e., repeated
observations at identical inputs. Replication offers a glimpse at pure
simulation variance, which is valuable for detecting a weak signal in high
noise settings. Replication also holds the potential for computational savings
through pre-averaging of repeated observations.
It becomes doubly essential when the noise level varies in the input space.
Although there are many ways to embellish the classical GP setup for
heteroskedastic modeling, e.g., through choices of the covariance kernel, few
acknowledge computational considerations.  In fact, many exacerbate the
problem.  A notable exception is {\em stochastic kriging}
\citep[SK,][]{ankennman:nelson:staum:2010} which leverages replication for
thriftier computation in low signal-to-noise regimes,
where it is crucial to distinguish intrinsic stochasticity from extrinsic
model uncertainty.  However, SK has several drawbacks. Inference for unknowns
is not based completely on the likelihood.  It has the crutch of {\em
requiring} (a minimal amount of) replication at each design site, which
limits its application. Finally, the modeling and extrapolation of
input-dependent noise is a secondary consideration, as opposed to one which is
fully integrated into the joint inference of all unknowns.

\blu{Our contributions address these limitations from both computational and
methodological perspectives. On the computational side, we expand upon a
so-called Woodbury identity to reduce computational complexity of inference
and prediction under replication: from the (log) likelihood of all parameters
and its derivatives, to the classical kriging equations. We are not the first
to utilize the Woodbury ``trick`` with GPs \citep[see,
e.g.,][]{opsomer:etal:1999,Banerjee2008,ng:yin:2012}, but we believe we are
the first to realize its full potential under replication in both
homoskedastic and heteroskedastic modeling setups.  We provide proofs of new
results, and alternate derivations leading to simpler results for quantities
first developed in previous papers, including for SK. For example, we
establish---as identities---results for the efficiency of estimators
leveraging replicate-averaged quantities, which previously only
held in expectation and under fixed parameterization.}

On the methodological end, we further depart from SK and borrow a
heteroskedastic modeling idea from the machine learning literature
\citep{goldberg:williams:bishop:1998} by constructing an apparatus to jointly
infer a spatial field of simulation variances (a noise-field)  along with the
mean-field GP. The structure of our proposed method is most similar to the one
described in \citet{kersting:etal:2007} who turn a ``hard expectation
maximization (EM)'' problem, with expectation replacing the expensive Markov
Chain Monte Carlo (MCMC) averaging over latent variables, into a pure
maximization problem. Here, we show how the Woodbury trick provides two-fold
computational and inferential savings with replicated designs: once for
conventional GP calculations arising as a subroutine in the wider
heteroskedastic model, and twice by reducing the number of latent variables
used to described the noise field.  We go on to illustrate how we may obtain
inference and prediction in the spirit of SK, but via full likelihood-based
inference and without requiring minimal replication at each design site.

A limitation of the \citeauthor{kersting:etal:2007} pure maximization approach
is a lack of smoothness in the estimated latents. We therefore propose to
interject an explicit smoothing {\em back} into the objective function being
optimized. This smoother derives from a GP prior on the latent (log) variance
process, and allows one to explicitly model and optimize the spatial
correlation of the intrinsic stochasticity. It serves to integrate noise-field
and mean-field in contrast to the SK approach of separate empirical
variance calculations on replicates, followed by independent auxiliary
smoothing for out-of-sample prediction. We implement the smoothing in such a
way that full derivatives are still available, for the latents as well as the
coupled GP hyperparameters, so that the whole inferential problem can be
solved by deploying an off-the-shelf library routine.  Crucially, our
smoothing mechanism does not bias the predictor at the
stationary point solving the likelihood equations, compared to an un-smoothed
alternative.  Rather, it has an annealing effect on the likelihood,
and makes the solver easier to initialize, both accelerating its convergence.

The remainder of the paper is organized as follows.  In Section \ref{sec:gp}
we review relevant GP details, SK, and latent noise variables.  Section
\ref{sec:woodbury} outlines the Woodbury trick for decomposing the covariance
when replication is present, with application to inference and prediction
schemes under the GP.  Practical heteroskedastic modeling is introduced in
Section \ref{sec:het}, with the Woodbury decomposition offering the double
benefit of fewer latent variables and faster matrix decompositions.  An
empirical comparison on a cascade of alternatives, on toy and real data, is
entertained in Section \ref{sec:empirical}.  We conclude with a brief
discussion in Section \ref{sec:discuss}.

\section{GPs under replication and heteroskedasticity}
\label{sec:gp}

Standard kriging builds a surrogate model of an unknown function $f:
\R^d \to \R$ given a set of noisy output observations $\vecY = (y_1, \dots,
y_N)^\top$ at design locations $\vecX = (\x_1, \dots, \x_N)^\top$. This is
achieved by putting a so-called Gaussian process (GP) prior on $f$,
characterized by a mean function and covariance function $k: \R^d \times
\R^d \to \R$. We follow the simplifying assumption in the computer experiments
literature in using a mean zero GP, which shifts all of the modeling effort to
the covariance structure. The covariance or kernel $k$ is positive
definite, with parameterized families such as the Gaussian or Mat\'ern being
typical choices.

Observation model is $y(\x_i) = f(\x_i) + \varepsilon_i$,
$\varepsilon_i \sim \mN(0, r(\x_i))$. In the typical homoskedastic case
$r(\x) = \tau^2$ is constant, but we anticipate heteroskedastic models as
well. In this setup, the modeling framework just described is equivalent to
writing $Y \sim \mathcal{N}_N(\bm{0}, \KN + \SN)$, where $\KN$ is the $N
\times N$ matrix with $ij$ coordinates $k(\x_i, \x_j)$, and
$\SN=\mathrm{Diag}(r(\x_1),\ldots, r(\x_N))$ is the noise matrix.
\blu{Notice that $\x_i$ is a $d \times 1$ vector and the $\varepsilon_i$'s are i.i.d..}

Given a form for $k(\cdot, \cdot)$, multivariate normal (MVN)
conditional identities provide a predictive distribution at site $\x$:
$Y(\x) | \vecY$, which is Gaussian with parameters
\begin{align*}
\mu(\x) &= \bE(Y(\x)| \vecY) =  \veck(\x)^\top (\KN
+ \SN)^{-1} \vecY, \; \mbox{ where } \veck(\x) = (k(\x, \x_1), \dots, k(\x, \x_N))^\top;
 \\
 \sigma^2(\x) &= \bV(Y(\x)| \vecY) =
k(\x, \x) + r(\x) - \veck(\x)^\top (\KN + \SN)^{-1} \veck(\x).
\end{align*}

The kernel $k$ has hyperparameters that must be estimated.
Among a variety of methods, many are based on the likelihood, which is simply a MVN density.
Most applications involve stationary kernels, $k(\x, \x') = \nu c(\x - \x'; \boldsymbol{\theta})$, i.e., \blu{$\KN = \nu \CN$}, with $\nu$ being the process variance and $\boldsymbol{\theta}$ being additional hyperparameters of the correlation function $c$.
 After relabeling $\KN + \SN = \nu(\CN + \LaN)$, first order optimality conditions provide a plug-in estimator for the common factor $\nu$: $\hat{\nu} = N^{-1}\vecY^\top (\CN +\LaN) ^{-1} \vecY$.
\blu{The log-likelihood conditional on $\hat{\nu}$} is then:
\begin{equation}
\log L =  -\frac{N}{2}\log (2 \pi) - \frac{N}{2} \log \hat{\nu}  - \frac{1}{2} \log |\CN + \LaN|  - \frac{N}{2}. \label{eq:ll}
\end{equation}
Observe that we must decompose \blu{(e.g., via Cholesky)} $\CN +
\LaN$ potentially multiple times in a maximization scheme, for inversion and
determinant computations, which requires $\Or(N^3)$ operations for the typical
choices of $c$.  This cubic computation severely limits the experiment size,
$N$, that can be modeled with GPs. Some work-arounds for large data include
approximate strategies \citep[e.g.,][]{Banerjee2008,haaland:qian:2012,kaufman:etal:2012,eidsvik2014estimation,gramacy:apley:2015} or a degree of tailoring to the
simulation mechanism or the input design
\citep{plumlee:2014,nychka:etal:2015}.

\textbf{Replicates:} When replication is present, some of the design sites are
repeated. This offers a potential computational advantage via switching
from the full-$N$ size of the data to the unique-$n$ number of unique design
locations. To make this precise, we introduce a special notation; note that
our setting is fully generic and nests the standard setting by allowing the
number of replicates to vary site-to-site, or be absent altogether.

Let $\xu_i$, $1= i, \dots, n$ represent the $n \ll N$ unique input
locations, and $y_i^{(j)}$ be the $j^\mathrm{th}$ out of $a_i \ge 1$ replicates,
i.e., $j=1,\dots, a_i$, observed at $\xu_i$, where $\sum\limits_{i = 1}^n a_i
= N$.  Then let $\vecYu = (\yu_1, \dots,
\yu_n)^\top$ collect averages of replicates, $\yu_i =
\frac{1}{a_i}\sum\limits_{j =1}^{a_i} y_i^{(j)}$. We now develop a map from
full $\KN, \SN$ matrices to their unique-$n$ counterparts. Without loss of
generality, assume that the data are ordered so that $\vecX = (\xu_1, \dots,
\xu_1, \dots
\xu_n)^\top$ where each input is repeated $a_i$ times, and where $\vecY$ is
stacked with observations on the $a_i$ replicates in the same order.  With
$\vecX$ composed in this way, we have $\vecX = \U \vecXu$, with $\U$ the
$N\times n$ block matrix $\U = \mathrm{Diag}(\ones_{a_1,1}, \dots,
\ones_{a_n,1}),$ where $\ones_{k,l}$ is $k \times l$ matrix filled with ones.
Similarly, $\KN = \U \Kn \U^\top$,
where $\Kn = (k(\xu_i, \xu_j))_{1 \leq i, j, \leq n}$ while $\U^\top \SN \U = \An \Sn$
where $\Sn = \Diag(r(\xu_1), \dots, r(\xu_n))$ and $\An = \Diag(a_1,
\dots, a_n)$. This decomposition is the basis for the Woodbury identities exploited in
Section \ref{sec:woodbury}. Henceforth, we utilize this notation with $n$ and
$N$ subscripts highlighting the size of matrices and vectors.

\subsection{Stochastic kriging}
\label{sec:stok}

A hint for the potential gains available with replication, while addressing
heteroskedasticity, appears in
\citet{ankennman:nelson:staum:2010}  who show that
the unique-$n$ predictive equations
\begin{align*}
\mu_n(\x) &= \veckn(\x)^\top (\Kn + {\An^{-1} \Sn})^{-1} \vecYu
\quad \mbox{ where } \quad\veckn(\x) = (k(\x, \xu_1), \dots, k(\x, \xu_n))^\top, \\
\sigma_n^2(\x) &= k(\x, \x) + r(\x) - \veckn(\x)^\top (\Kn + \An^{-1} \Sn)^{-1} \veckn(\x)
\end{align*}
are unbiased and minimize mean-squared prediction error.
\blu{This result implies
that one can handle $N \gg n$ points in $\Or(n^3)$ time.}
A snag is that one can rarely presume to know the variance function $r(\x)$.
\citeauthor{ankennman:nelson:staum:2010}, however show that with a sensible
estimate
\begin{equation}\blu{
\widehat{\boldsymbol{\Sigma}}_n = \Diag(\hat{\sigma}^2_1,
\dots, \hat{\sigma}^2_n), \quad \mbox{ where } \quad
\hat{\sigma}^2_i = \frac{1}{a_i - 1} \sum \limits_{j = 1}^{a_i} (y_i^{(j)} - \yu_i)^2,}
\label{eq:sigmahat}
\end{equation}
in place of $\Sn$ the resulting $\mu_n(\x)$ is still unbiased with $a_i
\gg 1$ (they recommend $a_i \geq 10$).  The predictive variance $\sigma_n^2$
still requires an $r(\x)$, for which no observations are directly available
for estimation.  One could specify $r(\x) = 0$ and be satisfied with an
estimate of extrinsic variance,  i.e., filtering heterogeneous noise
\citep{roustant:ginsbourger:deville:2012}, however that would preclude any
applications requiring full uncertainty quantification.  Alternatively,
\citeauthor{ankennman:nelson:staum:2010}, proposed a second, separately
estimated, (no-noise) GP prior for $r$ trained on the $(\xu_i,
\hat{\sigma}_i^2)$ pairs to obtain a prediction for $r(\x)$ at new $\x$
locations.

Although this approach has much to recommend it, there are several notable
shortcomings.
\blu{One is the difference of treatment between hyperparameters. The noise
variances are obtained from the empirical variance, while any other
parameter, such as lengthscales $\boldsymbol{\theta}$ for $k(\cdot, \cdot)$,
are inferred based on the pre-averaged log-likelihood (requiring $\Or(n^3)$ for evaluation):}
\begin{align}\label{eq:log-lik-sk}
{\log \bar{L} } :=
-\frac{n}{2}\log(2 \pi) - \frac{1}{2} \log |\Kn + \An^{-1} \widehat{\boldsymbol{\Sigma}}_n
|  -\frac{1}{2} \vecYu^\top (\Kn + \An^{-1}
\widehat{\boldsymbol{\Sigma}}_n)^{-1} \vecYu.
\end{align}
While logical and thrifty, this choice of $\widehat{\boldsymbol{\Sigma}}_n$ leads to the
requirement of a minimal number of replicates, $a_i > 1$ for all $i$, meaning
that incorporating even a single observation without a second replicate
requires a new method (lest such valuable observation(s) be dropped from the
analysis).  Finally, the separate modeling of the mean-field $f(\cdot)$ and
its variance $r(\cdot)$ is inefficient. Ideally, a likelihood would be
developed for all unknowns jointly.

\subsection{Latent variable process}
\label{sec:latent}

\citet{goldberg:williams:bishop:1998} were the first to propose a joint model
for $f$ and $r$, coupling a GP on the mean with GP on latent log variance (to
ensure positivity) variables. Although ideal from a modeling perspective, and
not requiring replicated data, inference for the unknowns involved a
cumbersome MCMC. Since each MCMC iteration involves iterating over each of
$\Or(N)$ parameters, with acceptance criteria requiring $\Or(N^3)$
calculation, the method was effectively $\Or(TN^4)$ to obtain $T$ samples, a
severely limiting cost.

Several groups of authors subsequently introduced thriftier alternatives in a
similar spirit, essentially replacing MCMC with
maximization
\citep{kersting:etal:2007,quadrianto:etal:2009,lazaro-gredilla:tsitas:2011},
but the overall computational complexity of each iteration of search remained
the same.  An exception is the technique of
\cite{boukouvalas:cornford:2009} who expanded on the EM-inspired method of
\cite{kersting:etal:2007} to 
exploit computational savings that comes from replication in the design.  However that
framework has two drawbacks.  One is that their mechanism for leveraging of
replicates unnecessarily, as we show, {\em approximates} predictive and
inferential quantities compared to full data analogues. Another is that,
although the method is inspired by EM, the latent log
variances are maximized rather than averaged over, yielding a solution
which is not a smooth realization from a GP.
\cite{quadrianto:etal:2009} addressed that lack of smoothness by introducing a
penalization in the likelihood, but required a much harder optimization.

In what follows we ultimately borrow the modeling apparatus of
\citeauthor{goldberg:williams:bishop:1998} and the EM-inspired method of
\citeauthor{kersting:etal:2007} and \citeauthor{quadrianto:etal:2009},
 but there are two important and novel ingredients
that are crucial to the adhesive binding them together.  We show that the
Woodbury trick, when fully developed below, facilitates massive computational
savings when there is large-scale replication, achieving the same
computational order as
\citeauthor{boukouvalas:cornford:2009} and SK but without approximation. We
then introduce a smoothing step that achieves an EM-style averaging over
latent variance, yet the overall method remains in a pure maximizing framework
requiring no simulation or otherwise numerical integration.
Our goal is to show that it is possible to get the best of all worlds: fully
likelihood-based smoothed and joint inference of the latent variances
alongside the mean-field (mimicking \citeauthor{goldberg:williams:bishop:1998}
but with the $O(n^3)$ computational demands of SK).  The only downside is that
our point estimates do not convey the same full posterior uncertainty as an
MCMC would.

\section{Fast GP inference and prediction under replication}
\label{sec:woodbury}

We exploit the structure of replication in design for GPs with extensive use
of two well-known formulas, together comprising the Woodbury identity
\citep[e.g.,][]{harville:1997}:
\begin{align}
 (\D + \U\B\V) ^{-1} &= \D^{-1} - \D^{-1}\U(\B^{-1} + \V\D^{-1}\U)^{-1}\V\D^{-1}; \label{eq:wood} \\
 |\D + \U \B \V| &= |\B^{-1} + \V \D^{-1} \U| \times |\B| \times |\D|,
\label{eq:matdet}
\end{align}
where $\D$ and $\B$ are invertible matrices of size $N \times N$ and $n \times
n$ respectively, and $\U$ and $\V^\top$ are of size $N\times n$.
In our context of GP prediction and inference, under the generic covariance parameterization $\KN + \SN = \nu(\CN + \LaN)$, we take the matrix $\D = \SN = \nu \LaN$
in \eqref{eq:wood} as diagonal, e.g., $\nu \LaN = \tau^2 \I_N$,
and $\V = \U^\top$. \blu{Moreover, $\SN$ shares the low dimensional structure,
$\U^\top \SN \U = \An \Sn = \nu \An \Lan $, and note that $\U^\top \U = \An$.}
In combination, these observations allow efficient
computation of the inverse and determinant of $(\KN + \SN)$.
In fact, there is no need to ever build the full-$N$ matrices.

\begin{lemma}
We have the following full-$N$ to unique-$n$ identities for the GP prediction equations, conditional on hyperparameters.
\begin{align}
\veckN(\x)^\top (\KN + \SN)^{-1} \vecY & = \veccn(\x)^\top (\Cn + \Lan \An^{-1})^{-1} \vecYu;  \label{eq:wood-mean} \\
\veckN(\x)^\top (\KN + \SN)^{-1} \veckN(\x) &= \nu\veccn(\x)^\top (\Cn + \Lan \An^{-1})^{-1} \veccn(\x). \label{eq:wood-var}
\end{align}
\end{lemma}

Eq.~\eqref{eq:wood-mean} establishes that the GP predictive mean, calculated
on the average responses at replicates and with covariances calculated only at
the unique design sites, is indeed identical to the original predictive
equations built by overlooking the structure of replication.
Eq.~\eqref{eq:wood-var} reveals the same result for the predictive variance. A
proof of this lemma is in Appendix \ref{sec:repproof}.
These identities support
common practice, especially in the case of the predictive mean, of only
utilizing the $n$ observations in $\vecYu$ for prediction.
\citet{ankennman:nelson:staum:2010}, for example, show that this unique-$n$
shortcut, applied via SK with $\Lan =
\nu^{-1} \widehat{\boldsymbol{\Sigma}}_n$ is the best linear unbiased predictor (BLUP), after conditioning
on the hyperparameters in the system.  This is simply a consequence of the
unique-$n$ predictive equations being identical to their full-$N$ counterpart,
inheriting BLUP and any other properties.

Although aspects of the results above have been known for some time, if
perhaps not directly connected to the Woodbury formula, they were not (to our
knowledge) known to extend to the full likelihood. The lemma below establishes
that indeed they do.

\begin{lemma}\label{lem:log-Ln}
Let $\Un := \Cn + \An^{-1} \Lan$. Then we have the following unique-$n$
identity for the full-$N$ expression for the conditional log likelihood,
$\log L$ in Eq.~(\ref{eq:ll}).
\begin{align}
&& \log L  &=  \mbox{Const} - \frac{N}{2} \log \nuN - \frac{1}{2} \sum\limits_{i=1}^n \left[(a_i - 1)\log \lambda_i + \log a_i \right] - \frac{1}{2} \log |\Un| , \label{eq:woodlik} \\
\mbox{where} && \nuN &:= N^{-1} \left(\vecY^\top \LaN^{-1} \vecY - \vecYu^\top \An \Lan^{-1}\vecYu + \vecYu^\top \Un^{-1} \vecYu \right). \label{eq:nuN}
\end{align}
\end{lemma}

The proof of the lemma is based on the following key computations:
\begin{align}
\vecY^\top (\CN + \LaN)^{-1} \vecY &= \vecY^\top \LaN^{-1} \vecY - \vecYu^\top \An \Lan^{-1} \vecYu + \vecYu^\top (\Cn + \An^{-1} \Lan)^{-1} \vecYu;  \label{eq:wood-y2} \\
\log |\CN + \LaN|  &= \log |\Cn + \An^{-1}\Lan| + \sum\limits_{i=1}^n \left[(a_i - 1)\log \lambda_i + \log a_i \right]. \label{eq:wood-det}
\end{align}
Assuming the $n\times n$ matrices have already been decomposed, at $\Or(n^3)$
cost, the extra computational complexity is $\Or(N + n)$ for the
right-hand-side of \eqref{eq:wood-y2} and $\Or(n)$ for the right-hand-side of
\eqref{eq:wood-det}, respectively.  Both are  negligible
compared to $O(n^3)$.

It is instructive to compare (\ref{eq:woodlik}--\ref{eq:nuN}) to the
pre-averaged log likelihood \eqref{eq:log-lik-sk} used by SK.
In particular, observe that the expression in \eqref{eq:nuN} is different to
the one that would be obtained by estimating $\nu$ with unique-$n$
calculations based on $\vecYu$, which would give $\nun = n^{-1}\vecYu^\top
\Un^{-1} \vecYu$, an $\Or(n^2)$ calculation assuming pre-decomposed matrices.
However, our full data calculation via $\vecY$ above gives $\nuN =
N^{-1}(\vecY^\top \LaN^{-1} \vecY
- \vecYu^\top \An \Lan^{-1} \vecYu + n\nun)$. The extra term in front of $\nun$
is an important correction for the variance at replicates:
\begin{equation}
N^{-1}(\vecY^\top \LaN^{-1} \vecY - \vecYu\An \Lan^{-1}\vecYu) = N^{-1} \sum_{i=1}^n \frac{a_i}{\lambda_i} s_i^2, \label{eq:sumvar}
\end{equation}
where $s_i^2 = \frac{1}{a_i} \sum_{j=1}^{a_i} (y_i^{(j)} - \bar{y}_i)^2$, i.e., the bias
un-adjusted estimate of $\mathbb{V}\mathrm{ar}(Y(\x_i))$ based on 
$\{y_i^{(j)}\}_{j=1}^{a_i}$.  Therefore, observe that as the $a_i$ get large,
Eq.~(\ref{eq:sumvar}) converges to $ N^{-1} \sum \limits_{i = 1}^n
\frac{a_i}{\lambda_i} \bV(Y(\x_i))$. Note that computing
\eqref{eq:sumvar} is in $\Or(N + n^2)$ with pre-decomposed matrices.

Finally, we provide the derivative of the unique-$n$ log
 likelihood to aid in numerical optimization, revealing
a computational complexity in $\Or(N + n^3)$, or essentially $\Or(n^3)$:
\begin{align}
\frac{\partial \log L}{\partial \cdot} &= \frac{N}{2} \frac{\partial \left( \vecY^\top \LaN^{-1} \vecY - \vecYu \An \Lan^{-1} \vecYu + n\nun \right)}{\partial \cdot} \times \left( N\nuN \right)^{-1} \nonumber \\
&- \frac{1}{2} \sum\limits_{i=1}^n \left[(a_i - 1) \frac{\partial \log \lambda_i}{\partial \cdot} \right] - \frac{1}{2} \mathrm{tr}\!\left( \Un^{-1} \frac{\partial \Un}{\partial \cdot} \right). \label{eq:lln}
\end{align}
The calculations above are presented  generically so that they can be applied
both in homoskedastic and heteroskedastic settings. In a homoskedastic
setting, recall that we may take $\lambda_i := \lambda = \tau^2 / \nu$, and no further
modifications are necessary provided that we choose a form for the covariance
structure, generating $\C_n$ above, which can be differentiated in closed
form. In the heteroskedastic case, described in detail below, we propose a
scheme for estimating the latent $\lambda_i$ value
at  the $i^\mathrm{th}$ unique design location $\xu_i$.  \blu{Appendix
\ref{sec:empwood} provides an empirical illustration of the computational
savings that comes from utilizing these identities in a homoskedastic context,
and  in the presence a modest degree of replication\eqref{eq:log-lik-sk}.}

\section{Practical heteroskedastic modeling}
\label{sec:het}

Heteroskedastic GP modeling involves learning the diagonal matrix $\LaN$ (or
its unique-$n$ counterpart $\Lan$), allowing the $\lambda_i$ values to exhibit
heterogeneity, i.e., not all $\lambda_i = \tau^2/\nu$ as in the homoskedastic
case. Care is required when performing inference for such a high dimensional
parameter. The strategy of \citet{goldberg:williams:bishop:1998} involves
applying regularization in the form of a prior that enforces positivity and
encourages a smooth evolution over the input space. Inference is performed by
integrating over the posterior of these latent values with MCMC. This is a
monumental task when $N$ is large, primarily because mixing of the latents is
poor.  We show that it is possible to have far fewer latent variables via
$\Lan$ (recall that $ \U^\top \LaN \U =
\An \Lan$) when replication is present, via the Woodbury identities above, and
to achieve the effect of integration by maximizing a criterion involving
smoothed quantities.

\subsection{A new latent process formulation}
\label{sec:newlatent}

A straightforward approach to learning $\Lan$ comes by maximizing the log
likelihood (\ref{eq:woodlik}).  As we show later in Figure \ref{tab:motor}
(top), this results in over-fit.  As in many high-dimensional settings, it
helps to regularize.  Toward that end
\citeauthor{goldberg:williams:bishop:1998} suggest a GP prior for \blu{$\log
\lambda$.} This constrains the \blu{$\log \Lan$} to follow a MVN law, but
otherwise they are free to take on any values.  In particular, if the MVN
acknowledges that our glimpse at the noise process, via the observed
$(x,y)$-values, is itself noisy, then the $\hat{\lambda}_i$s that come out, as
may be obtained from a local maximum of the resulting penalized likelihood
(i.e., the joint posterior for mean-field and noise processes),  may not be
smooth. Guaranteeing smoothed values via maximizing, as an alternative to more
expensive numerical integration, requires an explicit smoothing maneuver in
the ``objective'' used to solve for latents and hyperparameters.

Toward that end we re-cast the $\lambda_1, \dots, \lambda_n$ of $\Lan$ as
derived quantities obtained via the predictive mean of a regularizing GP on
new latent variables $\delta_1,
\dots,\delta_n$, stored in diagonal $\Deltan$ for ease of
manipulation.  That is
\begin{equation}
 \log \Lan = \Cg (\Cg + g \An^{-1})^{-1} \Deltan =:  \Cg \Ug^{-1} \Deltan,  \label{eq:lambdan}
\end{equation}
where $\Cg$ generically denotes the correlation structure of this noise
process, whose nugget is $g$, and $\Deltan \sim \mathcal{N}_n(0, \nug (\Cg +
g\An^{-1}))$.
The appearance of the replication counts $\An$ in $\Ug = (\Cg + g \An^{-1})$
is matching the structure of $\Un$ in Section \ref{sec:woodbury}; see Lemma \ref{lem:log-Ln}
or Eq.~\eqref{eq:wood-y2}. Thus the right-hand-side of \eqref{eq:lambdan} can be viewed as a
smoother of the latent $\delta_i$'s, carried out in analogue to the GP
equations for the mean-field. This setup guarantees that every $\Deltan$, e.g., as considered by
a numerical solver maximizing a likelihood, leads to smooth and positive variances
$\Lan$. Although any smoother could be used to convert latent $\Deltan$ into
$\Lan$, a GP mimics the choice of post-hoc smoother in SK.  The difference
here is that the inferential stage acknowledges a desire for smoothed
predictions too.

Choosing a GP smoother has important implications for the
stationary point of the composite likelihood developed in Section
\ref{sec:inference} below, positioning smoothing as a computational device
that facilitates automatic initialization of the solver and an annealing of
the objective for improved numerical stability.
The parameter $g$ governs the degree of smoothing: when $g=0$ we have $\log \Lan =
\Deltan$, i.e., no smoothing;  alternatively if $\Cg = \I_n$ and $\Deltan =
\I_n$, then $\log \Lan = (g+1)^{-1}\I_n$ and we recover the homoskedastic setup of
Section \ref{sec:intro} with noise variance \blu{$\tau^2 = \nuN \exp(1/(g+1))$}.

To learn appropriate latent $\Deltan$ values (leaving
details for other hyperparameters to Section \ref{sec:inference}),
we take advantage of a simple chain rule for the log-likelihood
(\ref{eq:lln}).
\begin{align}
\frac{\partial \log L}{\partial \Deltan} &= \frac{\partial \Lan}{\partial \Deltan} \frac{\partial \log L}{\partial \Lan} = \Lan \Cg \Ug^{-1} \frac{\partial \log L}{\partial \Lan},
\label{eq:log} \\
\mbox{ where } \quad \frac{\partial \log L}{\partial \lambda_i} &= \frac{N}{2} \times \frac{\frac{ a_i s_i^2} {\lambda_i^2} + \frac{(\Un^{-1} \vecYu)^2_i}{a_i}}{\nuN} - \frac{a_i - 1}{2\lambda_i} - \frac{1}{2a_i} (\Un)_{i, i}^{-1}. \label{eq:latentlik}
\end{align}
With $\mathbf{S} = \Diag(s_1^2, \dots, s_n^2)$, we may summarize (\ref{eq:latentlik}) in a more convenient vectorized form as
\begin{equation}
\frac{\partial \log L}{\partial \Lan} = \frac{1}{2} \frac{\An \mathbf{S} \Lan^{-2} + \An^{-1} \Diag(\Un^{-1} \vecYu)^2}{\nuN} - \frac{\An - \I_n}{2} \Lan^{-1} - \frac{1}{2}\An^{-1} \Diag(\Un^{-1}). \label{eq:dlambda}
\end{equation}

Recall that $s_i^2 = \frac{1}{a_i} \sum_{j=1}^{a_i} (y_i^{(j)} -
\bar{y}_j)^2$.  Therefore, an
interpretation of (\ref{eq:latentlik}) is as extension of the SK
estimate $\hat{\sigma}_i^2$ at $\xu_i$.  In contrast with SK, observe that the
GP smoothing is precisely what facilitates implementation for small numbers of
replicates, even $a_i = 1$, in which case even though $s_i^2 = 0$, $y_i$ still
contributes to the local variance estimates via the rest of
Eq.~(\ref{eq:latentlik}). Moreover, that smoothing comes nonparametrically via
all latent $\delta_i$ variance variables.  In particular, note that
(\ref{eq:latentlik}) is not constant in $\delta_i$; in fact it depends on all
of $\Deltan$ via Eq.~(\ref{eq:lambdan}).

\begin{remark}
Smoothing may be entertained on other quantities, e.g., $
\Lan\nuN = \Cg \Ug^{-1} \widehat{\boldsymbol{\Sigma}}_n^2$ (presuming $a_i >
1$), resulting in smoothed moment-based variance estimates in the style of SK,
as advocated by in \citet{kaminski:2015} and \citet{Wang2016}. There may
similarly be scope for bypassing a latent GP noise process with the so-called
SiNK predictor \citep{lee:owen:2015} by taking $\log \Lan =
\rho(\vecXu)^{-1}
\Cg \Ug^{-1}\Deltan$ with $\rho(\x) = \sqrt{\hnug \mathbf{c}_{(g)}(\x)^\top \Ug^{-1} \mathbf{c}_{(g)}(\x)}$.
\end{remark}

\subsubsection*{An illustration}

\begin{figure}[ht!]
\centering
\includegraphics[scale = 0.5, trim = 0 65 30 50, clip = TRUE]{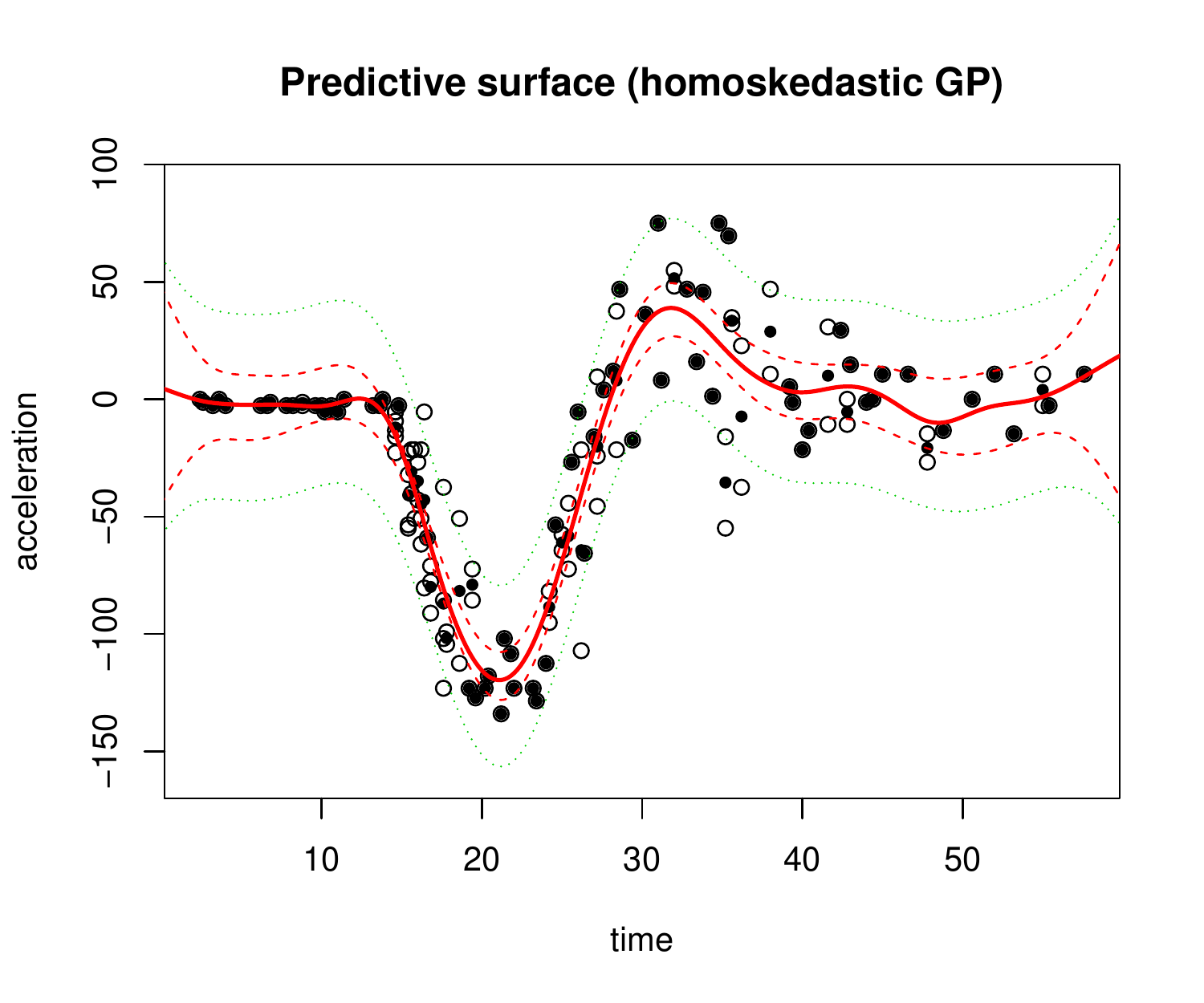}
\includegraphics[scale = 0.5, trim = 0 65 30 50, clip = TRUE]{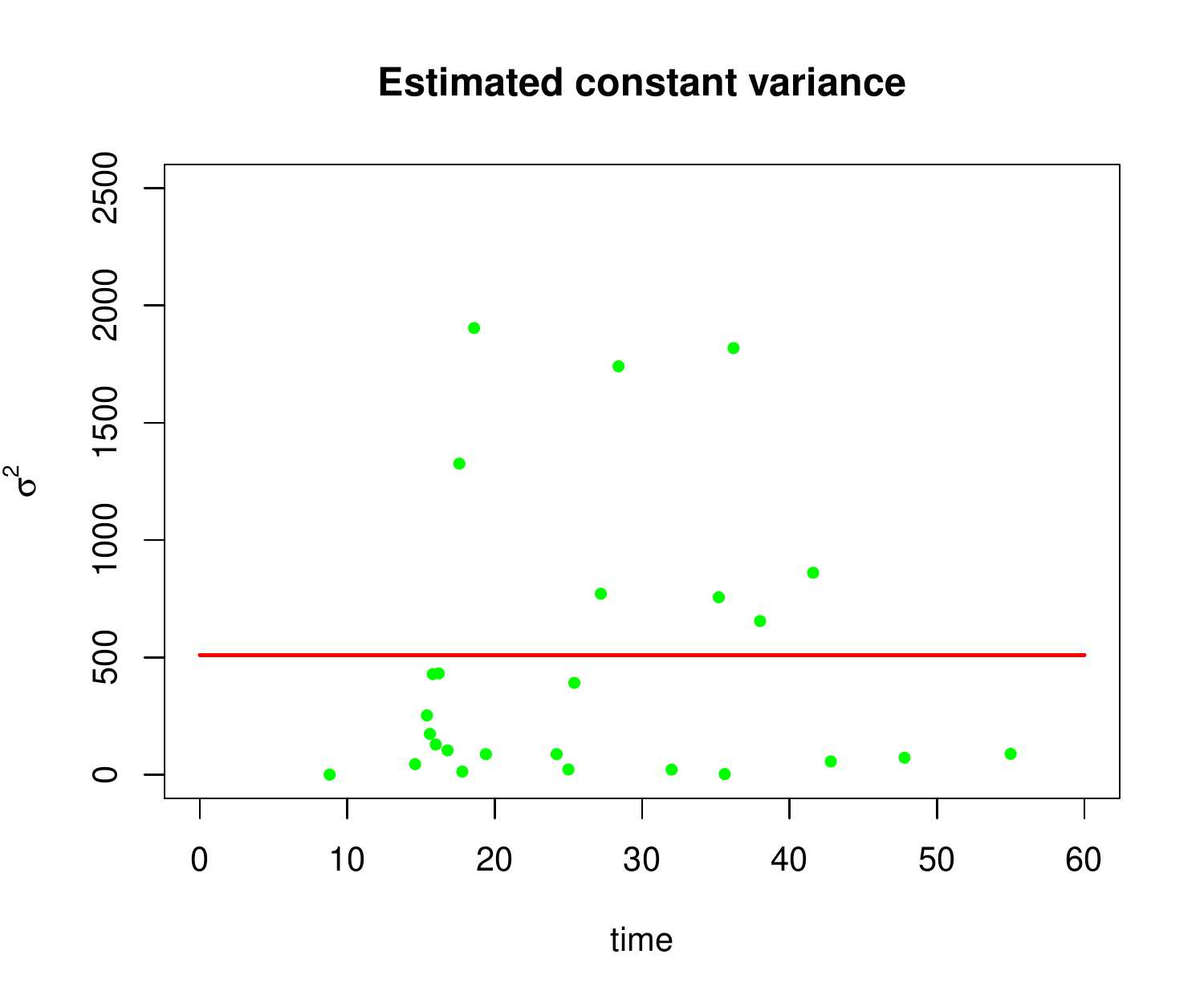}
\includegraphics[scale = 0.5, trim = 0 65 30 50, clip = TRUE]{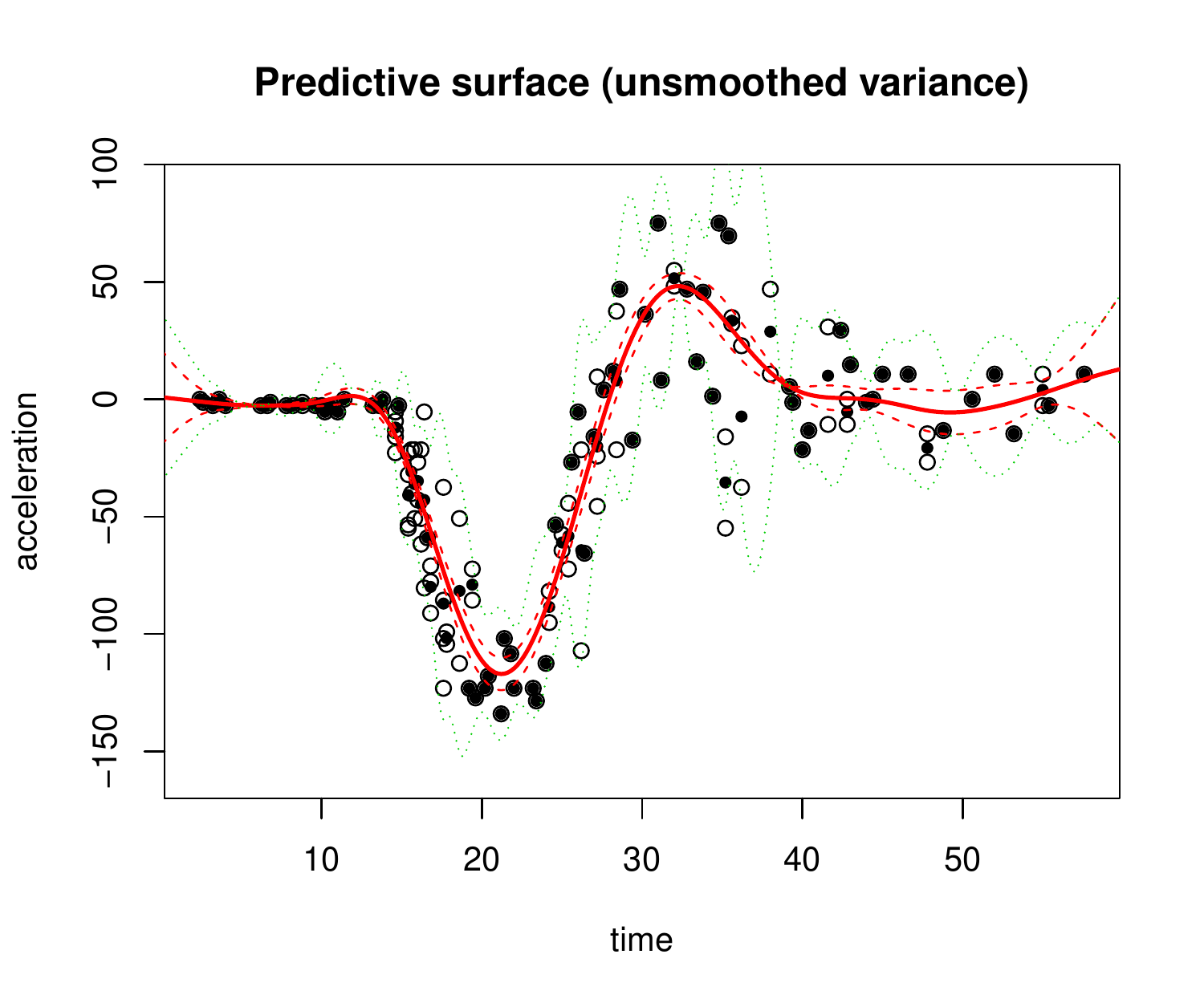}
\includegraphics[scale = 0.5, trim = 0 65 30 50, clip = TRUE]{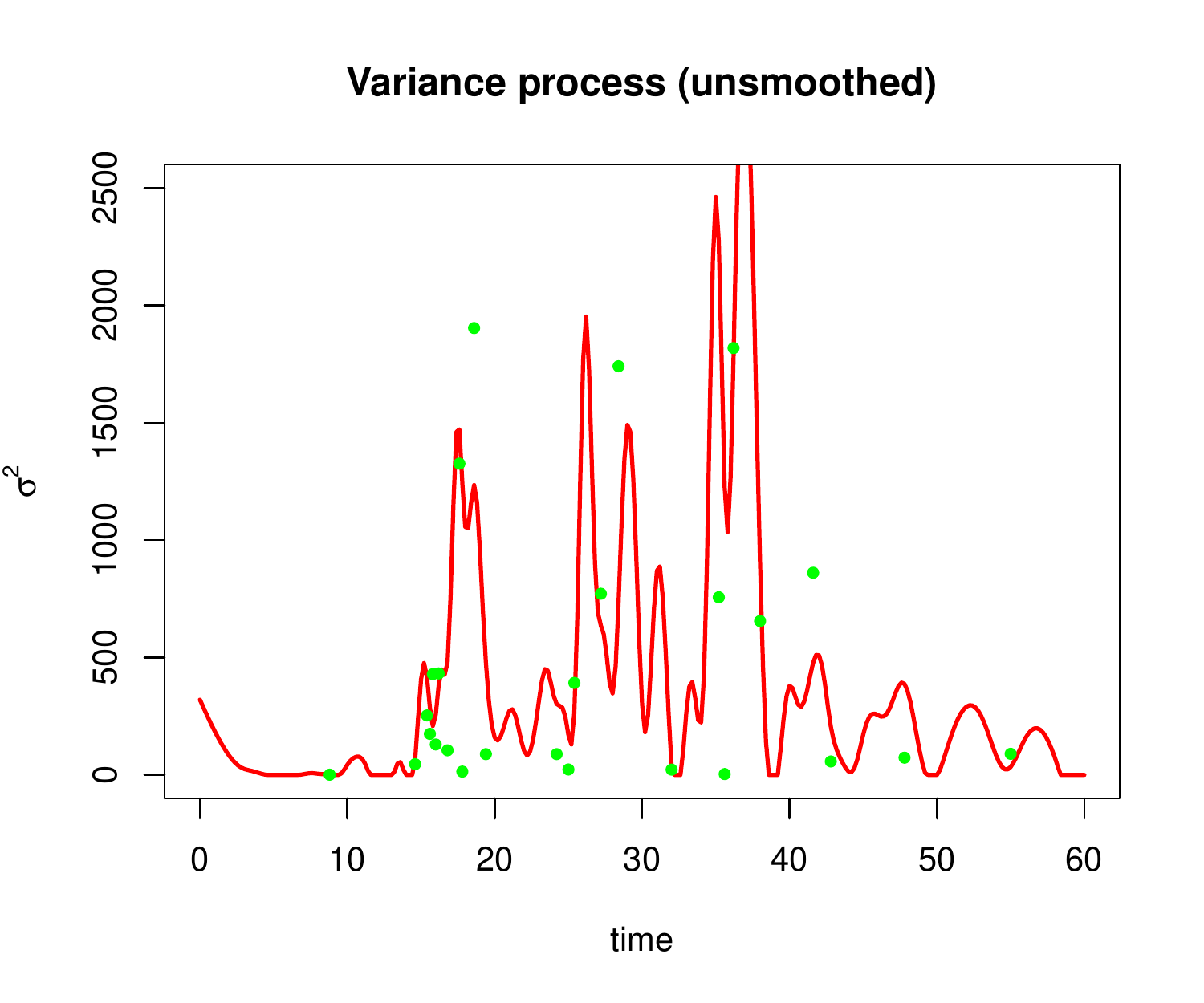}
\includegraphics[scale = 0.5, trim = 0 18 30 50, clip = TRUE]{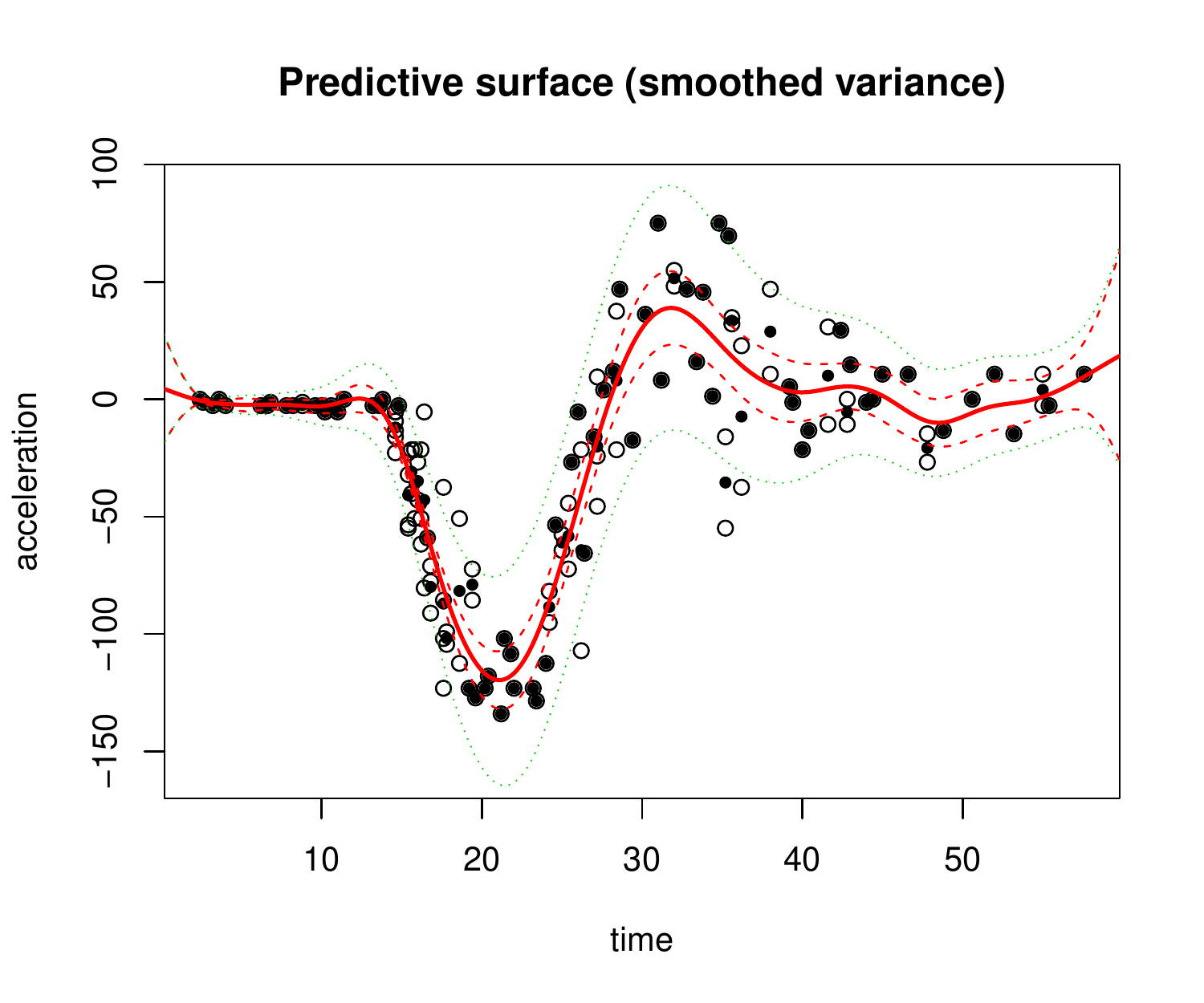}
\includegraphics[scale = 0.5, trim = 0 18 30 50, clip = TRUE]{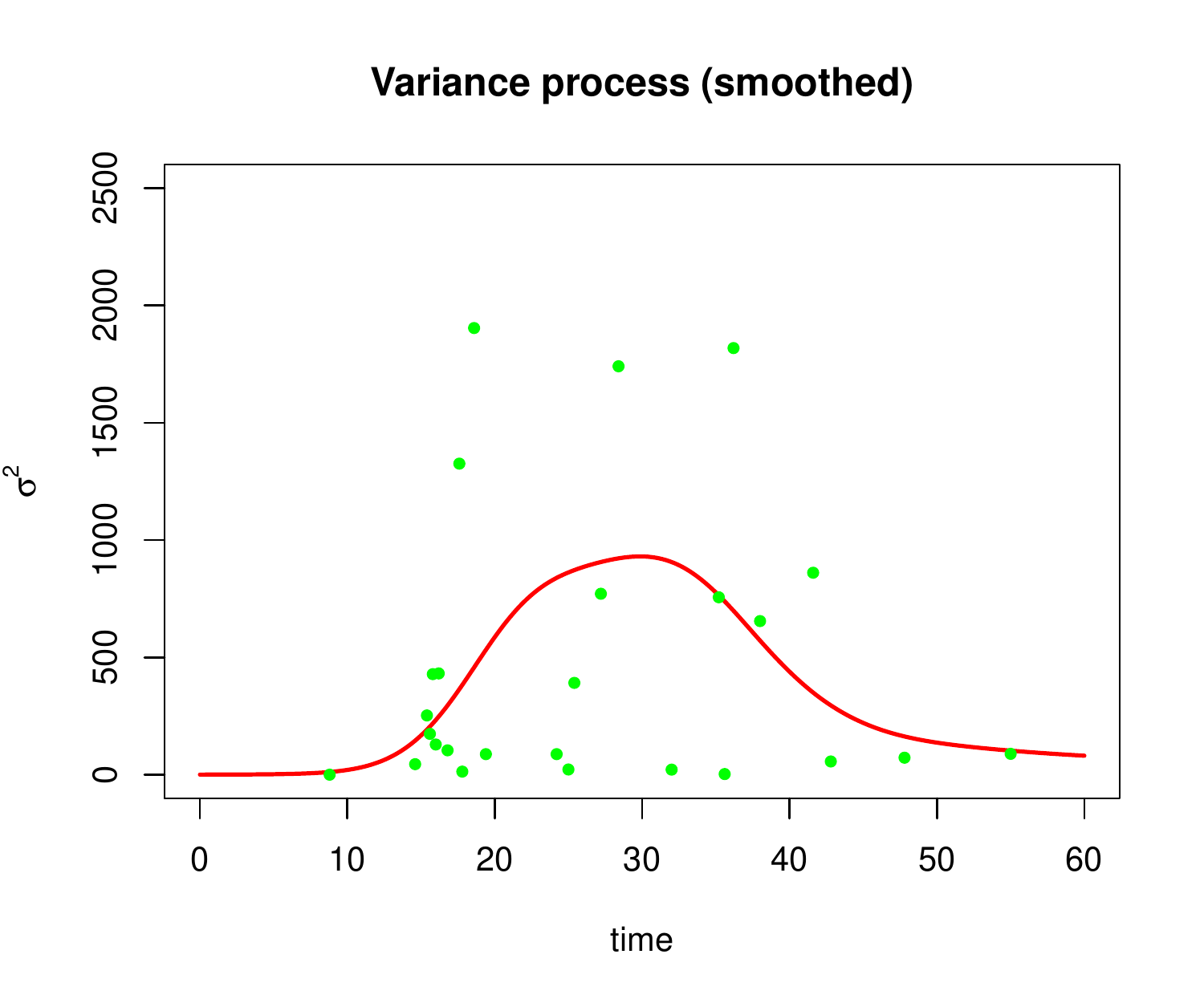}

\caption{\blu{Motorcycle example for constant noise (top), with (bottom) and without (center) smoothing the $\lambda_i$
values. {\em Left}: mean surface as a solid red line, with 95\%
confidence and prediction intervals in red dashed and green dotted
lines, respectively. Open circles are actual data points; averaged observations
$\bar{y}_i$ (for inputs with $a_i > 1$) are filled. {\em Right}: means of the variance
surface as red lines, empirical variance $\hat{\sigma}^2_i$ at replicates as green points.}}
\label{fig:motor2}
\end{figure}

The smoothing in Eq.~(\ref{eq:lambdan}) is particularly useful at input
locations where the degree of replication is low.  For an illustration,
consider the motorcycle accident data which is described in more detail in
Section \ref{sec:bench}. The SK method would not apply to this data, as inputs
are replicated at most six times, and often just once. Since $N=133$ and
$n=94$, the gains from the Woodbury trick are minor, yet there is value in
properly taking advantage of the few replicates available. The left-hand
panels of Figure \ref{fig:motor2} show the predictive surface(s) in terms of
predictive mean (solid red) and its 95\% confidence interval (dashed red) and
95\% prediction interval (dashed green).  The right-hand panels show the
estimated variance process.  \blu{The top row corresponds to a homoskedastic fit,
the middle row to the non-smoothed heteroskedastic case},
whereas the bottom row is based on
\eqref{eq:lambdan}.
Aesthetically, the bottom (smoothed variances) look better than the middle
(un-smoothed) alternative.
A more quantitative comparison, including alternatives from the literature,
 is provided in Section \ref{sec:bench}.

\subsection{Simultaneous optimization of hyperparameters and latents}
\label{sec:inference}

Smoothing (\ref{eq:lambdan}) is key to sensible evolution of the noise
process over the input space, but perhaps the most important feature of our
framework is how it lends itself to direct and tractable inference via
likelihood, without auxiliary calculation or simulation. Computational
complexity is linked to the degree of replication---$\Or(n^3)$ is much better
than $\Or(N^3)$ if $N \gg n$---however we remind the reader that no minimal
degree of replication is required.

The full set of hyperparameters are determined by the correlation structure(s)
of the two GPs, as well as the latent $\Deltan$. We denote by
$\boldsymbol{\theta}$ the lengthscales of the mean-field GP, generating the $n
\times n$ matrix $\Cn$. Similarly, we use $\boldsymbol{\phi}$ for the
lengthscales in $\Cg$, with the degree of smoothing \eqref{eq:lambdan}
determined by $g$. Inference requires choosing only values of these
parameters, $\{\boldsymbol{\theta}, \boldsymbol{\phi}, \Deltan, g\}$, because
conditional on those, the scales $\nu$ and $\nu_g$ of both processes have
plug-in MLEs: $\nuN$ in Eq.~(\ref{eq:nuN}) and analogously $\hnug = n^{-1}
\Deltan^\top \Ug^{-1} \Deltan$.  Therefore plugging in $\nuN$ and $\hnug$
yields the following concentrated joint log likelihood $\log \tilde{L}$:
\begin{align}
\log \tilde L =~&   - \frac{N}{2} \log \nuN  \nonumber - \frac{1}{2} \sum\limits_{i=1}^n \left[(a_i - 1)\log \lambda_i + \log a_i \right] - \frac{1}{2} \log |\Un|  \\
& - \frac{n}{2} \log \hnug  - \frac{1}{2} \log |\Ug| + \mbox{Const}, \label{eq:jllik}
\end{align}
where the top line above is the mean-field component, identical to
 (\ref{eq:woodlik}), and the bottom one is the analog for the latent variance
 process, all up to an additive constant.

Although seemingly close to the penalization used in
\cite{quadrianto:etal:2009}, a key difference here lies in the use of $g$,
analogous to a homogeneous noise on the variance process.  Smoothing, via $g$,
has an annealing effect on the optimization objective and, as Lemma
\ref{lem:g} establishes, can have no adverse effect on the final solution. Our
own preliminary empirical work (not shown) echoes results of
\cite{lazaro-gredilla:tsitas:2011} who demonstrate the value of smoothing in
this context as a guard against overfit.

\begin{lemma}
The objective $\log \tilde L$ in Eq.~(\ref{eq:jllik}) is maximized when
$\Deltan = \log \Lan$ and $g = 0$.
\label{lem:g}
\end{lemma}
The proof is left in Appendix \ref{ap:lemg}, however some discussion may be
beneficial.  This result says that smoothing (\ref{eq:lambdan}) is
redundant under the GP prior for $\Deltan$: you'll get a smooth answer anyway,
if you find the global maximum.
If we initialize $\Deltan$ at a non-smooth setting, say via an empirical
estimate of variance derived from residual sums of squares from an initial GP
fit, an (initially) non-zero $g$-value will compensate and automatically
generate a spatially smooth $\Lan$. As the solver progresses we can signal
convergence---to a smoothed $\hat{\boldsymbol{\Delta}}_n$---if $\hat{g}$ is
(numerically) zero.  Or, if on a tight computational budget, we can stop the
solver early, and rely on $\log \hat{\boldsymbol{\Lambda}}_n$ values being
smooth, but close to their $\hat{\boldsymbol{\Delta}}_n$ counterparts.

Inference for all unknowns may be carried out by a Newton-like
scheme since the gradient of (\ref{eq:jllik}) is available in closed form.
Its components may be derived generically through
Eqs.~(\ref{eq:lln}--\ref{eq:log}), i.e., without separating out by
hyperparameter, however we detail them below for
convenience. For each component $\theta_k$ of the mean-field lengthscale
$\boldsymbol{\theta}$ defining $\C_n$, we have
\begin{equation}
\frac{\partial \log \tilde L}{\partial \theta_k} = \frac{1}{2 \hat{\nu}_N} \vecYu^\top \Un^{-1} \frac{ \partial  \Cn }{\partial \theta_k} \Un^{-1} \vecYu
- \frac{1}{2} \mathrm{tr}\!\left( \Un^{-1} \frac{\partial \Cn}{\partial \theta_k} \right). \label{eq:dtheta}
\end{equation}
The rest of the components of the gradient depend on the derivative of the
 smoothed $\Lan$, i.e., $\frac{\partial \tilde L}{\partial \Lan}$ in
 (\ref{eq:dlambda}).   For the latent variance parameters $\delta_i$ in
 $\Deltan$, we have
\begin{align}
\frac{\partial \log \tilde L}{\partial \Deltan} &= - \frac{\Ug^{-1} \Deltan}{\hnug} +
\Cg \Ug^{-1} \Lan \times \frac{\partial \tilde L}{\partial \Lan}. \label{eq:ddelta} 
\end{align}
For each component $\phi_{k}$ of the lengthscale of the noise process, $\boldsymbol{\phi}$, we have
\begin{align}
\frac{\partial \log \tilde L}{\partial \phi_{k}} &= \left[ \frac{\partial \Cg}{\partial \phi_k} -\Cg \Ug^{-1}\frac{\partial \Cg}{\partial \phi_k} \right] \Ug^{-1}\Deltan \Lan \times \frac{\partial \tilde L}{\partial \Lan} \label{eq:dph} \\
& \qquad + \frac{1}{2\hnug}\Deltan^\top \Ug^{-1} \frac{\partial \Cg}{\partial \phi_k} \Ug^{-1}\Deltan
- \mathrm{tr}\!\left( \Ug^{-1} \frac{\partial \Cg}{\partial \phi_k} \right). \nonumber
\end{align}
And finally, for the noise-smoothing nugget parameter $g$ we have
\begin{align}
\frac{\partial \log \tilde L}{\partial g} &=  -\Cg \Ug^{-1} \An^{-1}\Ug^{-1} \Deltan \Lan \times \frac{\partial \tilde L}{\partial \Lan} \label{eq:dg} \\
& \qquad + \frac{1}{2\hnug}\Deltan^\top \Ug^{-1} \An^{-1} \Ug^{-1}\Deltan
- \mathrm{tr}\!\left(\An^{-1} \Ug^{-1}\right). \nonumber
\end{align}

\subsection{Implementation details}
\label{sec:implement}

Our implementation is in {\sf R} \citep{R}, \blu{available in the package
{\tt hetGP}} \cite{Binois2017}, essentially consists of feeding an objective, i.e., a coding
of the joint log-likelihood (\ref{eq:jllik}), and its gradient
(\ref{eq:dtheta}--\ref{eq:dg}), into the {\tt optim} library with
\verb|method="lbfgsb"|. 
A demonstration of the library, using the motorcycle data from Sections
\ref{sec:newlatent} \& \ref{sec:bench}, is provided in Appendix \ref{sec:motohet}. The
implementation is lean on storage and avoids slow {\tt for} loops in {\sf R}.
In particular vectors, not matrices, are used to store diagonals like
$\Deltan$, $\An$, $\Lan$, etc.  Traces of non-diagonal matrices are calculated
via compiled {\tt C++} loops, as are a few other calculations that would
otherwise have been cumbersome in {\sf R}. We have found that a thoughtful
initialization of parameters, and in some cases a sensible restriction of
their values (i.e., somewhat limiting the parameter space), results in a
faster and more reliable convergence on a tight computing budget. E.g., no
restarts are required.

Specifically, we find it useful to deploy a priming stage, wherein a single
homoskedastic GP (with the Woodbury trick) is fit to the data, in order to set
initial values for the latent variables and their GP hyperparameterization.
Initial $\delta_i$'s are derived from the logarithm of residual sums
of squares calculations on the fitted values obtained from the homoskedastic
GP, whose estimated lengthscales determine the initial $\boldsymbol{\theta}$
and $\boldsymbol{\phi}$ values,
{ $\delta_{i}^0 = \log \left( 1/a_i \sum_{j=1}^{a_i}(\mu(\x_i) - y_i^{(j)})^2 \right)$}.
An initial value for $g$ is obtained via a
separate, independent homoskedastic GP fit to the resulting collection of
$\Deltan$, which may also be utilized to refine the starting
$\boldsymbol{\phi}$ values. Our software includes the option of restricting the search for
lengthscales to values satisfying $\boldsymbol{\phi} \geq
\boldsymbol{\theta}$, taken component-wise. This has the effect of forcing the noise
process to evolve more slowly in space than the mean-field process does. A
thrifty approximate solution that avoids deploying a solver handling
constraints involves lower-bounding $\boldsymbol{\phi}$ based on values
of $\boldsymbol{\theta}$ obtained from the initial homogeneous fit. We include
a further option restricting $\boldsymbol{\phi} = k \boldsymbol{\theta}$ where the scalar $k
\geq 1$ is inferred in lieu of $\boldsymbol{\phi}$ directly, which is beneficial in higher dimensions.

Latent variable estimates returned by the optimizer can provide a good
indication of the nature of convergence.  Any $\delta_j$  at the boundary of
the \verb|lbfgsb| search region either indicates that the bounding box
provided is too narrow, the initialization is poor, or that $g$ needs to be
larger because the $\Lan$ values are being under-smoothed.  Although we have
iteratively re-engineered our ``automatic default'' initializations and other
settings of the software to mitigate such issues, we find that in fact such
adjustments usually have small impact on the overall performance of the
method. \blu{Numerical experiments backing the proposed initialization scheme are
given in Appendix \ref{ap:opt}.}

\blu{After the solver has converged we conclude the procedure with comparison of
the unpenalized log-likelihood, i.e., the top part of Eq.~\eqref{eq:jllik}, to
one obtained from an optimized homoskedastic fit. The latter may be higher
if the optimization process was unsuccessful, however typically the
explanation is that there was not enough data to justify an input-dependent
noise structure.  In that case we return the superior homoskedastic fit.}

\begin{remark}
\blu{It is worth noting that working directly with latent variances $\Lan$, rather
than log variances $\log \Lan$, might be appropriate if it is unreasonable to
model the variance evolution multiplicatively.  The expressions provided above
(\ref{eq:dtheta}--\ref{eq:dg}) simplify somewhat compared to our favored
logged version.  However, a zero-mean GP prior on the (un-exponentiated)
$\Lan$ could yield negative predictions, necessitating thresholding. We
suggest augmenting with a non-zero constant mean term, i.e., $\hat{\mu}_{(g)}
=
\bm{1}^\top
\Ug^{-1} \Deltan \left(\bm{1}^\top \Ug^{-1} \bm{1} \right)^{-1}$ as a
potential means of reducing the amount of thresholding required.}
\end{remark}

\section{Empirical comparison}
\label{sec:empirical}

In what follows we consider three examples, starting with a simple expository
one from the literature, followed by two challenging ``real data'' computer
model applications from inventory optimization and epidemiology.

\subsection{Benchmark data}
\label{sec:bench}

One of the classical examples illustrating heteroskedasticity is the
motorcycle accident data. It consists of $N=133$ accelerations with respect to
time in simulated crashes, $n=94$ of which are measured at unique time inputs.
Hence, this is a somewhat unfavorable example as far as computational
advantage goes (via Woodbury), however we note that a method like SK would
clearly not apply.  Figure \ref{fig:motor2} provides a visualization of our
performance on this data. For the experimental work reported below we use the
heteroskedastic fit from the bottom row of that figure, corresponding to
smoothed (log) latent variance parameters under full MLE inference for all
unknowns, denoted WHGP. \blu{See Appendix \ref{sec:motohet} for an implementation
via {\tt mleHetGP} in our {\tt hetGP} library.} We also include a homoskedastic GP
comparator via our Woodbury computational enhancements, using {\tt mleHomGP},
abbreviated WGP.

Table \ref{tab:motor} summarizes our results alongside
ones of other methods in the literature, as extracted from
\citet{lazaro-gredilla:tsitas:2011}: another homoskedastic GP without Woodbury (labeled GP), a MAP alternative to
the \citet{goldberg:williams:bishop:1998} heteroskedastic GP (labeled MAPHGP) from
\cite{kersting:etal:2007,quadrianto:etal:2009} and a variational alternative (VHGP)
from  \cite{lazaro-gredilla:tsitas:2011}.
\begin{table}[htpb]
\centering
\begin{tabular}{l|c|c|c|c|c}
     & WGP       & WHGP & GP ($\star$) & MAPHGP ($\star$) & VHGP ($\star$) \\ \hline
NMSE & 0.28 $\pm$ 0.21     & 0.28 $\pm$ 0.21 & 0.26 $\pm$ 0.18 & 0.26 $\pm$ 0.17  & 0.26 $\pm$ 0.17 \\ \hline
NLPD & 4.59 $\pm$ 0.25 & 4.26 $\pm$ 0.31 & 4.59 $\pm$ 0.22 & 4.32 $\pm$ 0.60 & 4.32 $\pm$ 0.30
\end{tabular}
\caption{Average Normalized MSE and Negative Log-Probability Density (NLPD)
$\pm 1\  \mathrm{sd}$ on the motorcycle data over 300 runs for our methods
(WGP and WHGP) compared to those ($\star$) reported in
\cite{lazaro-gredilla:tsitas:2011}. Lower numbers are better.}
\label{tab:motor}
\end{table}
The experiment involves 300  random (90\%, 10\%) partitions into training and
testing data, with performance on the testing data averaged over all
partitions.  Although we generally prefer a Mat\'ern correlation structure, we
used a Gaussian kernel for this comparison in order to remain consistent with the
other comparators. The metrics used, which are derived from the references
above to avoid duplicating the efforts in those studies, are Normalized Mean
Squared Error (NMSE) and Negative Log-Probability Density (NLPD), the latter
being, up to a constant, the negative of the proper scoring rule we prefer in
our next example. For details see \citet{lazaro-gredilla:tsitas:2011}. The
take-away message from the table is that our proposed methods are comparable,
in terms of accuracy, to these comparators.
\blu{We report that our WHGP method took less than one second to perform 100 {\tt
optim} iterations.}

\subsection{Assemble to order}
\label{sec:ato}

Here we consider the so-called ``assemble-to-order'' (ATO) problem first
introduced by \citet{hong:nelson:2006}. At  its heart it is an optimization (or
reinforcement learning) problem, however here we simply treat it as a response
surface to be learned.  Although the signal-to-noise ratio is relatively high,
ATO simulations are known to be heteroskedastic, e.g., as illustrated by the
documentation for the {\sf MATLAB} library we utilized for the simulations
\citep{xie:frazier:chick:2012}. The problem centers around inventory
management for a company that manufactures $m$ products. Products are built
from base parts, called items, some of which are ``key'' in that the product
cannot be built without them.  If a request comes in for a product which is
missing one or more key items, a replenishment order is executed, and is filled
after a random period.  Holding items in inventory is expensive, so there is a
balance to be struck between the cost of maintaining inventory, and revenue
which can only be realized if the inventory is sufficient to fulfill demand for
products.

The inventory costs, product revenue, makeup of products (their
items), product demand stream and distribution of the random replenishment
time of items, together comprise  the problem definition. Here we use the
canonical specification of \citet{hong:nelson:2006} involving five products
built from eight items.  The input space is a target stock vector $b \in
\{0,1,\dots,20\}^8$ for the item inventories, and the ATO simulator provides
a Monte Carlo estimate of expected daily profit under that regime.

For the experiment reported on below we evaluated ATO on a Latin Hypercube
sample of size $2000$ in the discrete 8-dimensional space, sampling ten
replicates at each site.  We then embarked on fifty Monte Carlo repetitions of
an out-of-sample predictive comparison obtained by partitioning the data into
training and testing sets in the following way.  We randomly chose half
($n=1000$) of the sites as training sites, and at each site collected $a_i
\sim \mathrm{Unif}\{1,\ldots,10\}$ training responses among the ten replicates
available.  In this way, the average size of the training data was
$N=5000$.  The ten replicates at the remaining $n=1000$ testing sites comprise
our out-of-sample testing set (for $10000$ total).  We further retain the
random number of unchosen (i.e., also out-of-sample) training replicates (of
which there are $5000$ on average) as further testing locations.

\begin{figure}[ht!]
\centering
\vspace{-0.5cm}
\includegraphics[scale=0.6]{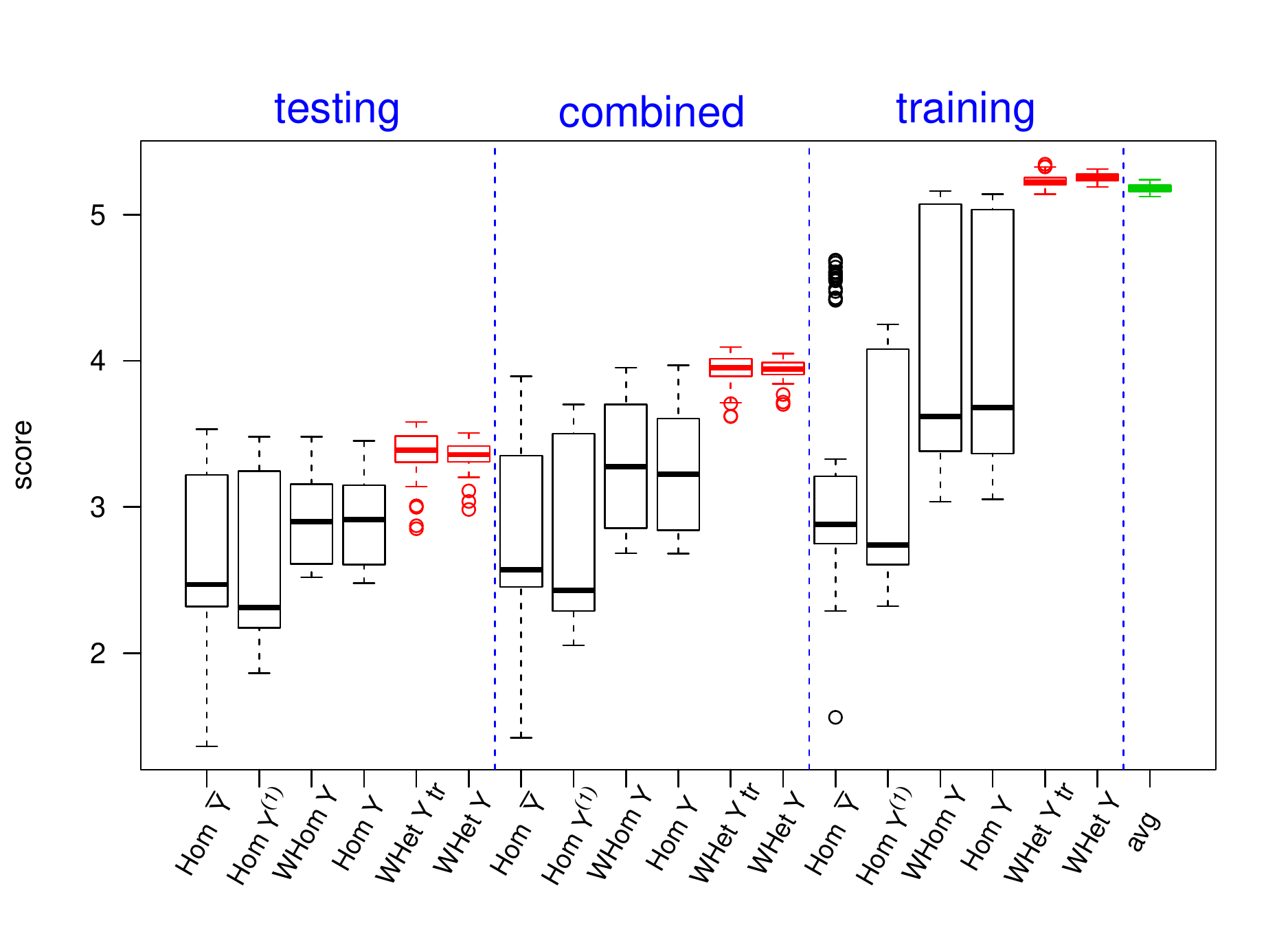}
\includegraphics[scale=0.6]{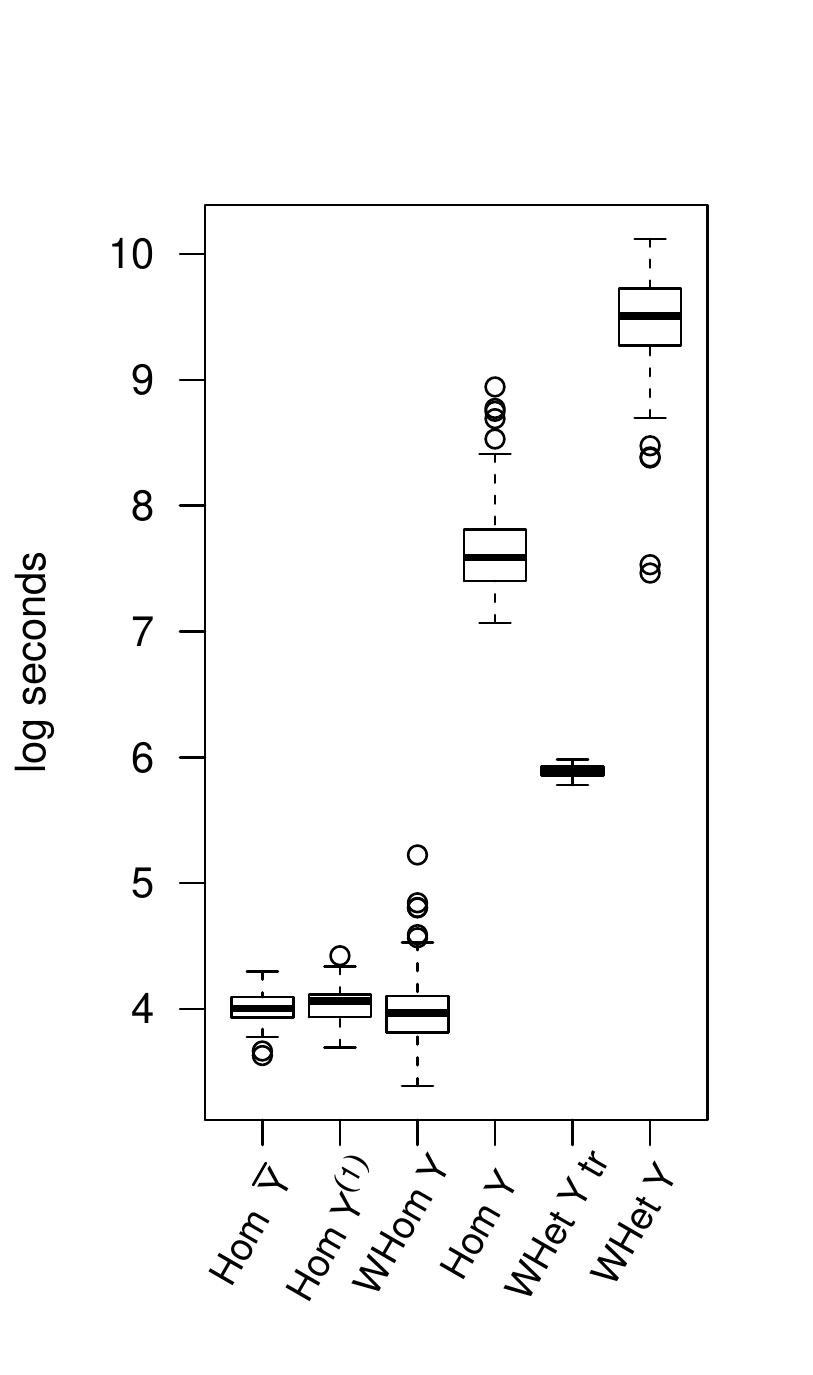}

\vspace{-0.5cm}
\caption{The {\em right} panel shows scores (higher is better) on ATO data comparing
homoskedastic partial and full-data comparators, and heteroskedastic
comparators (in red).  Sets are partitioned into pure out-of-sample
``testing'', out-of sample ``training'' at the input-design locations, and
``combined''.  The green {\em right}-most boxplot corresponds to the empirical
mean and variance comparator at the ``training'' sites.  The {\em right} panel
shows computing times in log seconds.}
\label{f:ato}
\end{figure}

The {\em left} panel of Figure \ref{f:ato} summarizes the distribution of
scores combining predictions of the mean and variance together via the proper
scoring rule provided by Eq.~(27) of \citet{gneiting:raftery:2007}.  The plot
is divided into four segments, and in the first three the same six comparators
are entertained. The first segment tallies scores on the testing design
locations, the third segment on the (out-of-sample) training locations, and
the second segment combines those two sets.  The final segment shows the
purely in-sample performance of a $(\bar{y}_i, s^2_i)$ estimator (``avg'') for
benchmarking purposes. The six comparators, from left to right, are ``Hom
$\bar{Y}$'' (a homoskedastic GP trained on $n=1000$ pairs ($\bar{Y},
\bar{X}$), i.e., otherwise discarding the replicates), ``Hom $Y^{(1)}$'' (a
homoskedastic GP trained on $n=1000$ runs comprised of only one (namely first) replicate in
the training data at each site), ``WHom $Y$'' (a homoskedastic GP on the full
training runs using the Woodbury trick), ``Hom $Y$'' (a homoskedastic GP on
the full training runs without the Woodbury trick), ``WHet $Y$ tr'' (a
\emph{truncated} heteroskedastic GP on the full training runs, using the
Woodbury trick but stopping after 100 {\tt optim} iterations), and finally
``WHet $Y$'', which is the same without limiting {\tt optim}.  To ease
viewing, the heteroskedastic comparators are colored in red.

Although there is some nuance in the relative comparison of the boxplots, the
take-home message is that the heteroskedastic comparators are the clear
winners. The fact that on the (out-of-sample) training runs the
heteroskedastic comparator is better than ``avg'' indicates that smoothed
variances are as crucial for good performance at design sites as it is
elsewhere (where smoothing is essential).  The wall-clock times of the relative
comparators are summarized in the {\em right} panel of the figure.  Observe
that ``WHom $Y$'' is much faster than ``Hom $Y$'' with a speed-up of about
$40\times$. This illustrates the value of the Woodbury identities; recall that here
$n\approx N/5$. Also, comparing ``WHet $Y$ tr'' and ``WHet $Y$'', we see that there is
little benefit from running {\tt optim} to convergence, which often consumes
the vast bulk of the computing time (a difference of over $50\times$ in the
figure), yet yields minimal improvement in the score for these comparators.

\subsection{Epidemics management}
\label{sec:SIR}

Our last example concerns an optimization/learning problem that arises in
the context of designing epidemics countermeasures
\citep{ludkovski:niemi:2010,merl:etal:2009}. Infectious disease outbreaks, such as
influenza, dengue fever and measles, spread stochastically through contact
between infected and susceptible individuals. A popular method to capture
outbreak uncertainty is based on constructing a stochastic compartmental
model that partitions the population into several epidemiological classes
and  prescribes the transition rates among these compartments. One of the
simplest versions is the stochastic SIR model that is based on Susceptible,
Infected and Recovered counts, with individuals undertaking $S \to I$ and $I
\to R$ moves. In a continuous-time formulation, the state $\mathbf{x}(t) :=
(S_t, I_t, R_t)$, $t\in \mathbb{R}_+$ is a  Markov chain taking values on
the simplex $\mathcal{X}_M = \{ (s,i,r) \in \mathbb{Z}_+^3, s+i+r \le M \}$
where $M$ is the total population size, here fixed for convenience. The SIR
system has two possible transition channels, with corresponding rates
\citep{hu2015sequential}
\begin{align}\label{eq:sir}
\left\{ \begin{aligned}
\text{Infection:}~ & S+I \to 2I  & &\text{with rate}\;\; \beta S_t I_t/M; \\
\text{Recovery:}~ & I \to R  & &\text{with rate}\;\; \gamma I_t. \\
\end{aligned}\right\}
\end{align}

Policy makers aim to dynamically enact countermeasures in order to mitigate
epidemic costs; a simple proxy for the latter is the total
number of individuals who are newly infected
$
f(\mathbf{x}) := \mathbb{E}[ S_0 - \lim_{T \to \infty} S_T | (S_0,I_0,R_0)
= \mathbf{x}] = \gamma \mathbb{E} \bigl[ \int_0^\infty I_t \,dt | \mathbf{x}
\bigr].
$
Due to the nonlinear transition rates, there is no closed-form expression
for the above expectation, however it can be readily estimated via Monte
Carlo by generating outbreak trajectories and averaging.
The final optimization step, which
we do not consider here, performs a cost-benefit analysis by comparing
expected costs under different policy regimes  (e.g., ``do-nothing'', ``public
information campaign'', ``vaccination campaign'').

The form of \eqref{eq:sir} induces a strong heteroskedasticity in the
simulations. For low infected counts $I \le 10$, the outbreak is likely to
die out quickly on its own, leading to low  expected infecteds and low
simulation variance. For intermediate values of $I_0$, there is a lot of
variability across scenarios, yielding high conditional variance
$r(\mathbf{x})$. However, if one starts with a lot of initial infecteds, the
dynamics become essentially deterministic due to the underlying law of large
numbers, leading to large costs but low simulation noise. The demarcation of
these regions is driven by initial susceptible count and the
transition rates $\beta,\gamma$ in \eqref{eq:sir}.

As proof of concept, we consider learning $f$ and $r$ for the case $\beta =
\gamma = 0.5$, $M = 2000$. The input space is restricted to $\{ (S,I):  S \in
\{1200,\ldots, 1800\}, I
\in \{0, \ldots, 200\} \}$.  Outside those ranges the outbreak behavior is
trivial. Inside the region, simulation noise ranges from $r(\mathbf{x}) = 0$
at $I_0=0$ to as high as $r(\mathbf{x}) = 86^2$. Figure 6 in
\cite{hu2015sequential} shows a variance surface generated via SK
using $a_i \equiv 100$ replicates at each input and an adaptive design.
Here we propose a more challenging
setup where the number of replicates is not fixed and includes small counts.
To this end, we generate a large training set with slightly more than two thousand
locations on a regular grid, with a thousand replications. Then we compare the
results given by SK and our method based on a random subset of $n=1000$
$\mathbf{x}_i$'s with a varying number of replicates. Specifically, 500 design sites
have only 5 replicates (250 with 10, 150 with 50 and 100 with 100,
respectively). To account for variance known to be zero
at $I = 0$, all design points on this domain boundary are artificially replicated 100 times,
so $N = \sum_i a_i = 24205$ on this specific instance.

\begin{figure}[ht!]
\centering%
	\begin{subfigure}[t]{0.32\textwidth}%
	\centering%
	\includegraphics[height=4.5cm]{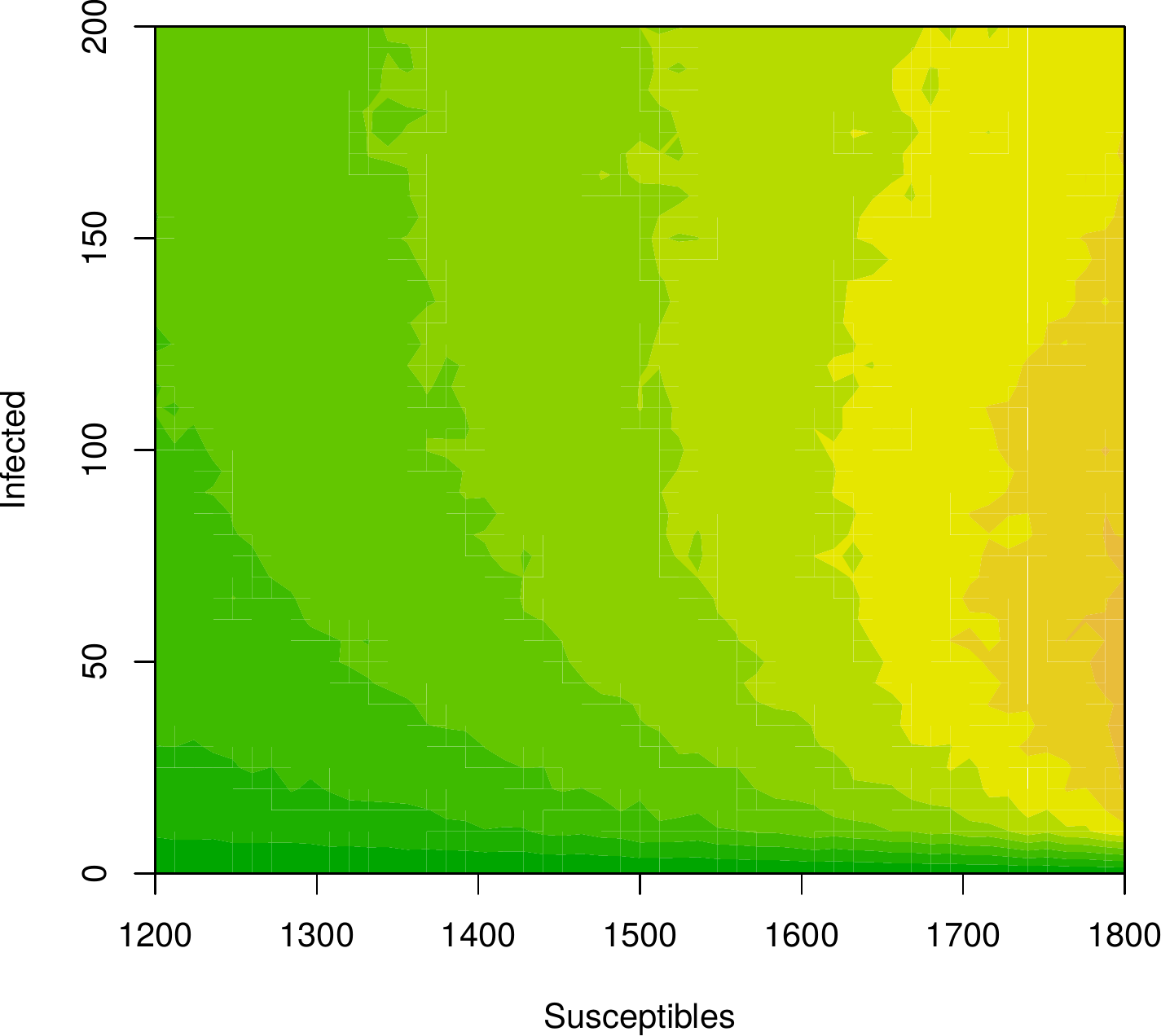}%
	\caption{\footnotesize Reference}%
	\end{subfigure}%
	\begin{subfigure}[t]{0.32\textwidth}%
	\centering%
	\includegraphics[height=4.5cm]{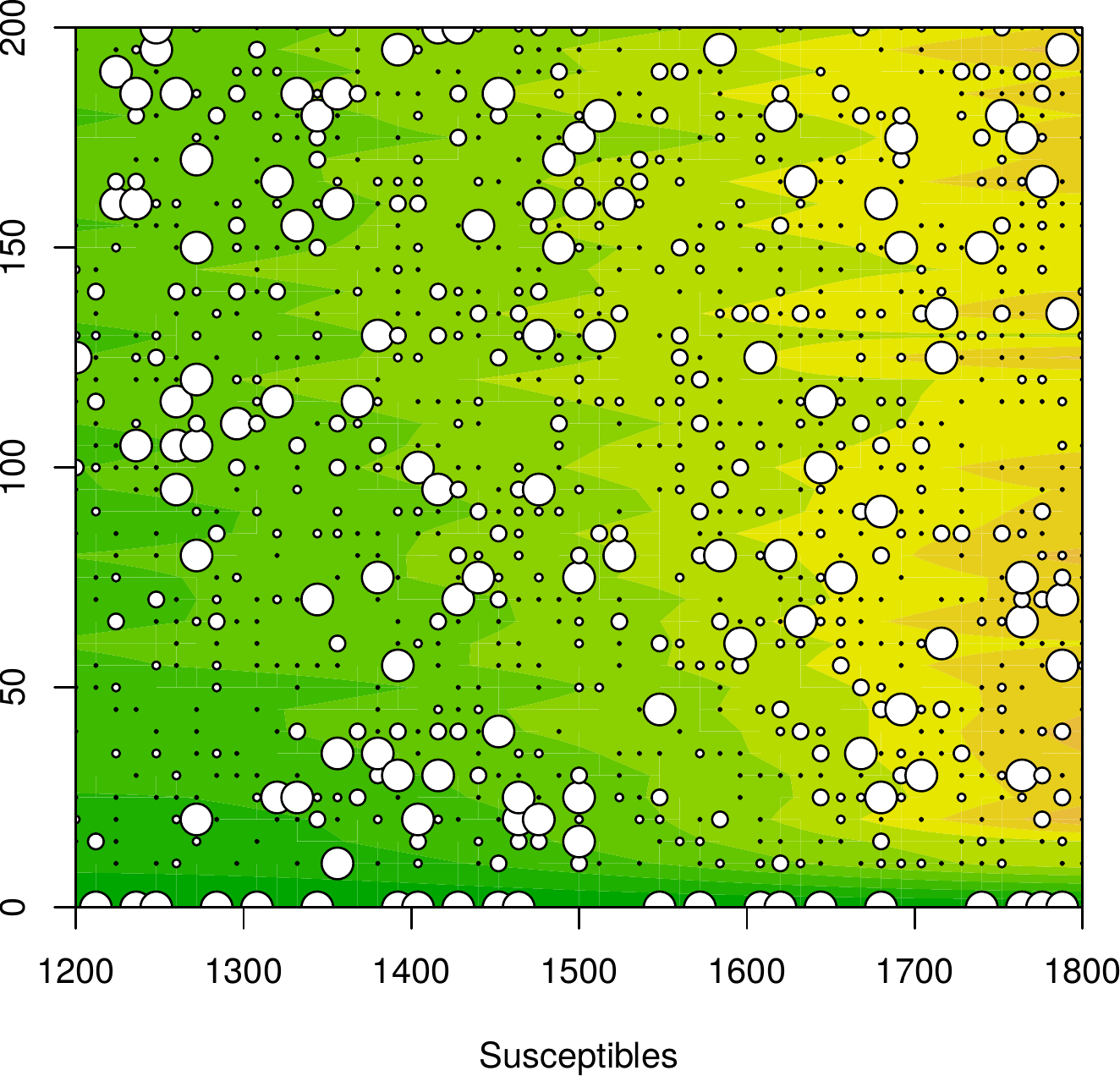}%
	\caption{\footnotesize Joint estimation}%
	\end{subfigure}
	\begin{subfigure}[t]{0.36\textwidth}%
	\centering%
	\includegraphics[height=4.5cm]{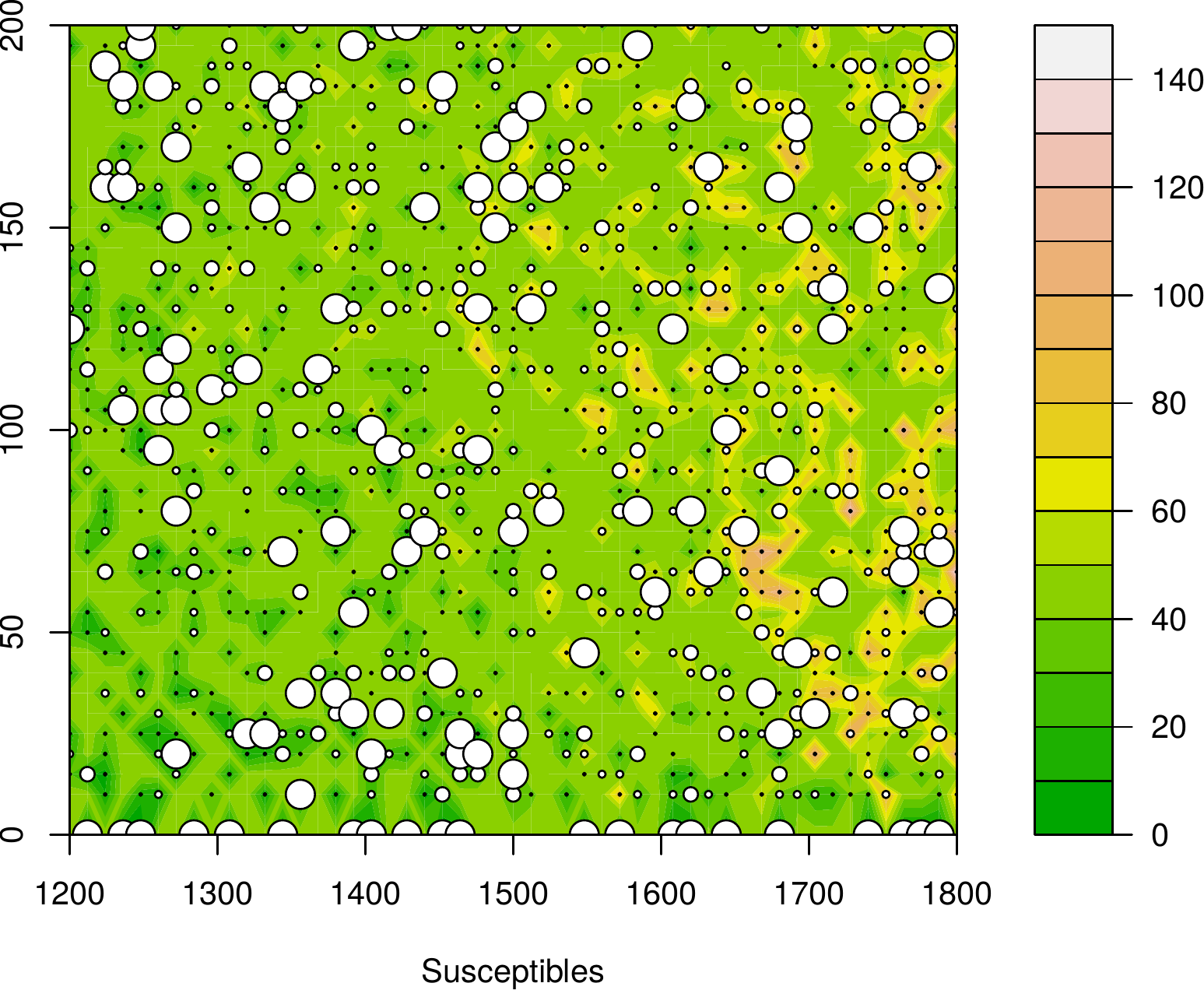}%
	\caption{\footnotesize Empirical (interpolated)}%
	\end{subfigure}%
	\caption{Estimated standard deviation surfaces for the epidemics example with several methods.
	Left: reference empirical standard deviation on all two thousand locations
	and all thousand replicates. Center and right: results based on one
	thousand locations with varying number of replications, depicted by points
	of increasing size (i.e., for 5, 10, 50 and 100 replicates). Center:
	results given from the joint estimation procedure. Right: interpolation of
	the empirical standard deviation at given $\mathbf{x}_i$'s. }
\label{fig:SIRVarSurfs}
\end{figure}

The counterpart of Figure 6 in \cite{hu2015sequential} is Figure
\ref{fig:SIRVarSurfs} here. Our proposed method recovers the key features of the
variance surface, with a significantly lower number of replicates. As a
comparison (see rightmost panel), an interpolating model directly fit onto the empirical
variances, as suggested in \cite{ankennman:nelson:staum:2010} is much less
accurate. It is then not surprising that SK is impacted by the inaccuracies
of the empirical mean and variance estimates.
\begin{figure}[ht!]
\centering
\vspace{-0.75cm}
\includegraphics[height=7.8cm]{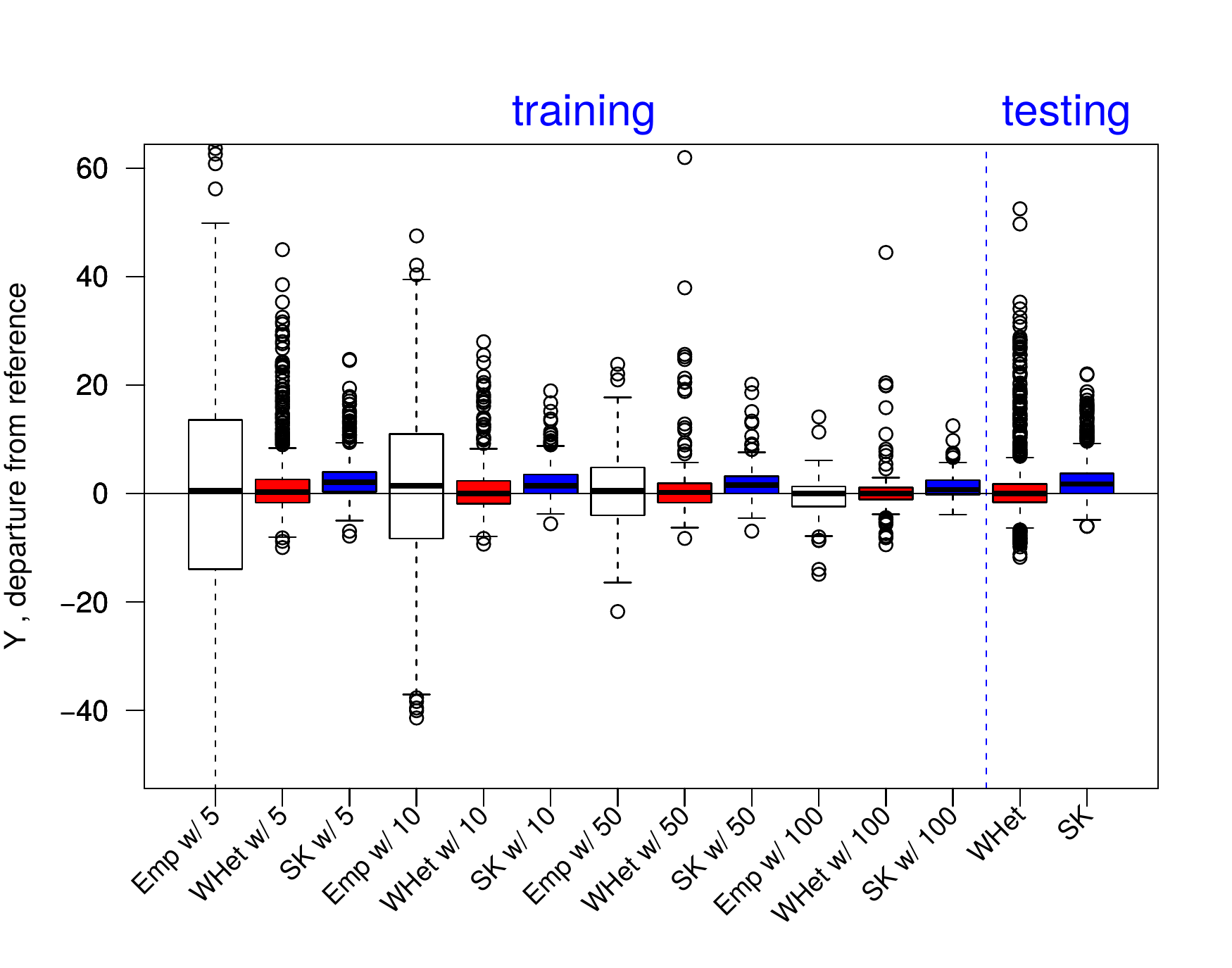}
\includegraphics[height=7.8cm]{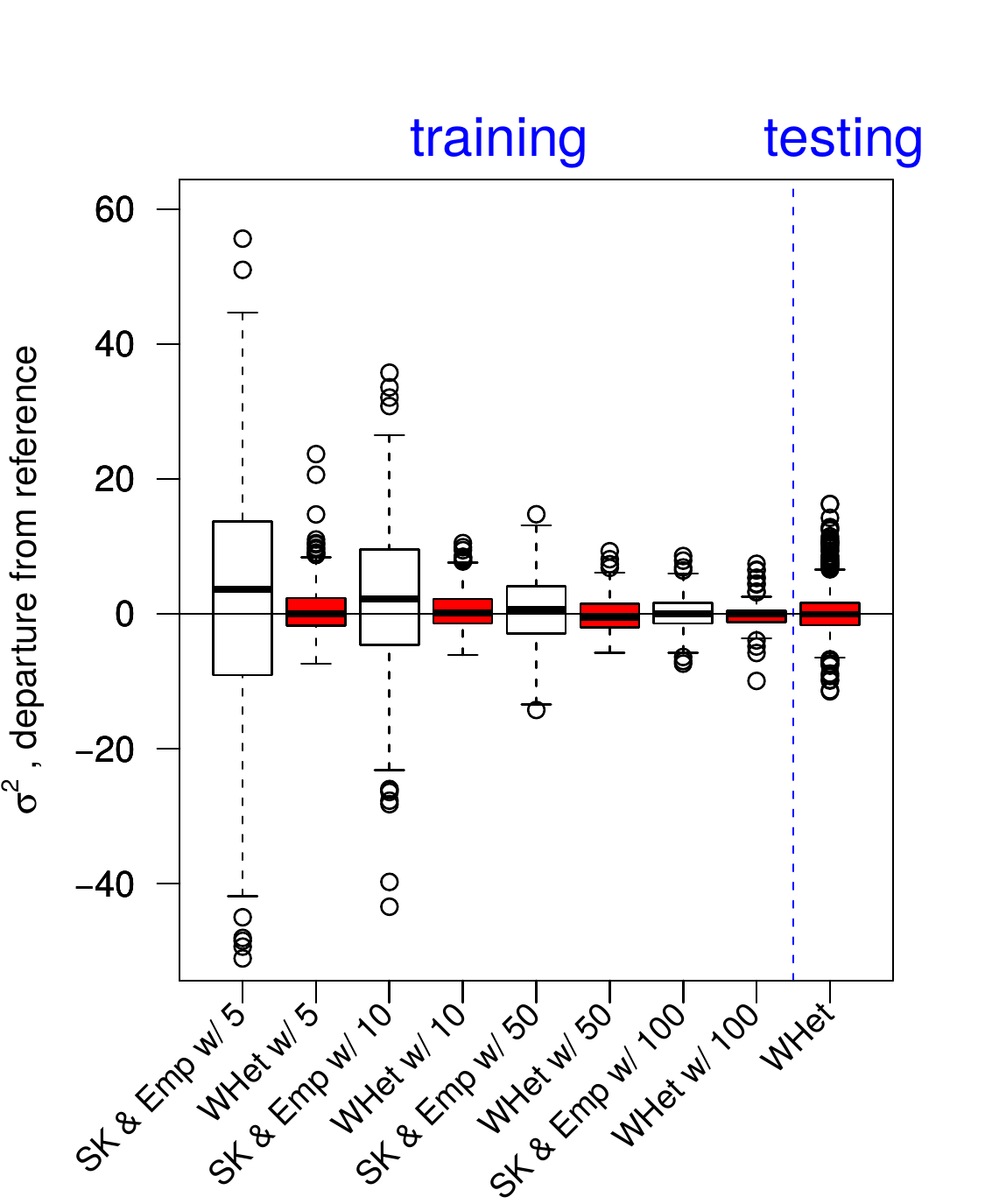}
\caption{Boxplot of difference between reference mean ({\em left} panel) and reference
standard deviation ({\em right} panel) obtained on all data with, on a subset
of locations and replicates, the empirical mean (Emp, white), the results
given by Stochastic Kriging (SK, blue) and those of our method (WHet, red).
Results are separated by the number of replicates in the training set (5 to 100),
as well as comparison on the testing set (extreme right). }
\label{fig:SIRMeanBoxes}
\end{figure}
In particular, we show in the
{\em left} panel of Figure~\ref{fig:SIRMeanBoxes} that the mean prediction for SK is off even at inputs
with many replicates ($a_i = 50$ or $a_i=100$), while our method benefits from the smoothing
of latent variables. In addition, we can perform the same analysis on the
predicted variance, showing again in the {\em right} panel of Figure~\ref{fig:SIRMeanBoxes} that our
estimates outperform the use of the empirical $\hat{\sigma}_i$'s, even for the largest
number of replicates. Observe that the empirical estimates are notably high-biased when $a_i < 50$.

\section{Discussion}
\label{sec:discuss}

We have enhanced heteroskedastic Gaussian process modeling by, effectively,
combining the strengths of canonical approaches in two disparate literatures,
namely machine learning (via \citeauthor{goldberg:williams:bishop:1998})~and
industrial design (via \citeauthor{ankennman:nelson:staum:2010}).  We
introduced two novel aspects, the Woodbury trick for computational shortcuts
(without approximation), and a smoothed latent process in lieu of expensive EM
and MCMC simulation-based alternatives to inference.  The result is an
inferential framework that upgrades modeling fidelity---input-dependent
noise---while retaining a computational complexity on par with the best
implementations of constant noise GP methods. Appendix \ref{ap:nonstat}
explores how our approach copes well with other types of non-stationarity,
despite the model mismatch.

\blu{In our survey of the literature we discovered a connection between the
Woodbury trick and an {\em approximate} method of circumventing big data
(homoskedastic) GP calculations via {\em pseudo inputs}
\citep{snelson:ghahramani:2005} and the so-called {\em predictive process}
\cite{Banerjee2008}. Those methods, rather than leveraging replicates,
approximates a big data spatial field with a smaller latent process: the
inputs, outputs, and parameters for which are jointly inferred in a unified
optimization (pseudo-inputs) or Bayesian (predictive process) framework.
Indeed the spirit of pseudo-input version is similar to ours, being almost
identical when restricting the latent process to the coordinates of the
unique-$n$ inputs $\vecXu$ in the homoskedastic modeling setup.
\citet{snelson:ghahramani:2006} and later \citet{kersting:etal:2007} proposed
heteroskedastic pseudo-inputs variants, with raw optimization of the latents
rather than via the smoothing process that we propose.  An advantage of free
estimation of the pseudo inputs, i.e., rather than fixing them to $\vecXu$ as
we do, is that additional computational savings can be realized in situations
where replication is infeasible or undesired, albeit by approximation and
assuming that the pseudo-inputs can be selected efficiently.}  We see
tremendous potential for further hybridization of our approach with the wider
literature on approximate (heteroskedastic) modeling of GP in big data
contexts.

\if0\blind
{
\subsubsection*{Acknowledgments}

All three authors are grateful for support from National Science Foundation grants
DMS-1521702 and DMS-1521743.  Many thanks to Peter Frazier for suggesting the ATO problem.
}\fi

\appendix

\section{Supporting replication}

\blu{The subsections below provide some of the details that go into lemmas
involving the Woodbury trick for efficient inference and prediction with
designs under replication, followed by an empirical illustration of the
potential computational savings using our library routines.}

\subsection{Derivations supporting lemmas in Section \ref{sec:woodbury}}
\label{sec:repproof}

Due to the structure of replicates, it follows that $\U^\top \U = \An$, $\An \veckn(\x) = \U^\top \veck(\x)$, $\U \veckn(\x) = \veck(\x)$. Moreover, $\An \vecYu = \U^\top \vecY$, $\U^\top \SN = \Sn \U^\top$, $\U \Sn = \SN \U$, $\An \Sn = \U^\top \SN \U$ (formulas with $\SN$ work with $\SN^{-1}$ as well).
Woodbury's identity gives (\ref{eq:wood}):
\begin{align*}
\veckn(\x)^\top &(\Kn + \An^{-1} \Sn)^{-1} \vecYu = \veckn(\x)^\top  \Sn^{-1} \An \vecYu - \veckn(\x)^\top \An \Sn^{-1}  (\Kn^{-1} + \An \Sn^{-1})^{-1} \Sn^{-1} \An \vecYu\\
&= \veckn(\x)^\top \Sn^{-1} \U^\top \vecY - \veck(\x)^\top \U \Sn^{-1}  (\Kn^{-1} + \An \Sn^{-1})^{-1} \Sn^{-1} \U^\top \vecY\\
&= \veckn(\x)^\top  \U^\top \SN^{-1} \vecY - \veck(\x)^\top  \SN^{-1} \U  (\Kn^{-1} + \An \Sn^{-1})^{-1} \U^\top \SN^{-1} \vecY\\
&= \veck(\x)^\top \SN^{-1} \vecY - \veck(\x)^\top  \SN^{-1} \U  (\Kn^{-1} + \U^\top \SN^{-1} \U)^{-1} \U^\top \SN^{-1} \vecY = \veck(\x)^\top (\K + \SN)^{-1} \vecY.
\end{align*}
The calculations are the same when replacing $\vecYu$ by $\veckn(\x)$ on the right.
On the other hand, having $\vecY$ on both sides yields:
\begin{align*}
\vecY^\top(\KN + \SN)^{-1}\vecY &= \vecY^\top \SN^{-1} \vecY - \vecY^\top \SN^{-1} \U(\Kn^{-1} + \U^\top \SN^{-1} \U)^{-1} \U^\top \SN^{-1} \vecY\\
&= \vecY^\top \SN^{-1} \vecY - \vecY^\top  \U \Sn^{-1}(\Kn^{-1} +\An \Sn^{-1})^{-1}  \Sn^{-1} \U^\top \vecY\\
&= \vecY^\top \SN^{-1} \vecY - \vecYu^\top \An \Sn^{-1} (\Kn^{-1} + \An \Sn^{-1})^{-1}\Sn^{-1} \An \vecYu\\
&= \vecY^\top \SN^{-1} \vecY - \vecYu^\top \An \Sn^{-1}\vecYu + \vecYu^\top (\Kn + \An^{-1} \Sn)^{-1} \vecYu.
\end{align*}
The determinant is similar:
\begin{align*}
\log |\K + \SN| &= \log|\Kn^{-1} +  \U^\top \SN^{-1} \U| + \log |\Kn| + \log |\SN|\\
&= \log|\Kn| + \log|\Kn^{-1} + \An \Sn^{-1}| +  \sum\limits_{i=1}^n a_i \log \Sigma_i \\
&= \log |\Kn + \An^{-1}\Sn|  + \sum\limits_{i=1}^n \left[(a_i - 1)\log \Sigma_i + \log a_i \right]\\
&= \log |\Kn + \An^{-1}\Sn|  + \log |\SN| - \log |\An^{-1} \Sn|.
\end{align*}

\subsection{Empirical demonstration of Woodbury speedups}
\label{sec:empwood}

\blu{The {\sf R} code below sets up a design in 2-d and evaluates the response as
$y(\x) = x_1 \exp\{-x_1^2 - x_2^2)$ \citep{Gramacy2007}, observed with
$\mathcal{N}(0, 0.01^2)$ noise.  Our example in Appendix \ref{ap:nonstat}
provides a visual [in Figure \ref{fig:nonstat2d}] of this response in a
slightly different context. The design has $n=100$ unique space-filling
locations, where their degree of replication is determined uniformly at random
in $\{1,\dots,50\}$ so that, on average, there are $N=2500$ total elements in
the design.}

{\singlespacing
\begin{verbatim}
R> library(lhs)
R> Xbar <- randomLHS(100, 2)
R> Xbar[,1] <- (Xbar[,1] - 0.5)*6 + 1
R> Xbar[,2] <- (Xbar[,2] - 0.5)*6 + 1
R> ytrue <- Xbar[,1] * exp(-Xbar[,1]^2 - Xbar[,2]^2)
R> a <- sample(1:50, 100, replace=TRUE)
R> N <- sum(a)
R> X <- matrix(NA, ncol=2, nrow=N)
R> y <- rep(NA, N)
R> ybar <- rep(NA, 100)
R> nfill <- 0
R> for(i in 1:100) {
+    X[(nfill+1):(nfill+a[i]),] <-
+       matrix(rep(Xbar[i,], a[i]), ncol=2, byrow=TRUE)
+    y[(nfill+1):(nfill+a[i])] <- ytrue[i] + rnorm(a[i], sd=0.01)
+    ybar[i] <- mean(y[(nfill+1):(nfill+a[i])])
+    nfill <- nfill + a[i]
+  }
\end{verbatim}
}

\noindent \blu{Below, we use the homoskedastic modeling function {\tt mleHomGP} from our {\tt
hetGP} package in two ways.  First, we use it as intended, with the Woodbury
trick, saving the the wall time.}

{\singlespacing
\begin{verbatim}
R> eps <- sqrt(.Machine$double.eps)
R> Lwr <- rep(eps,2)
R> Upr <- rep(10,2)
R> un.time <- system.time(un <- mleHomGP(list(X0=Xbar, Z0=ybar,
+   mult=a), y, Lwr, Upr))[3]
\end{verbatim}
}

\noindent \blu{Then, we use exactly the same function, but with arguments that
cause the Woodbury trick to be bypassed---essentially telling the
implementation that there are no replicates.}

{\singlespacing
\begin{verbatim}
R> fN.time <- system.time(fN <- mleHomGP(list(X0=X, Z0=y,
+   mult=rep(1,N)), y, Lwr, Upr))[3]
\end{verbatim}
}

\blu{First, lets compare times.}

{\singlespacing
\begin{verbatim}
R> c(fN=fN.time, un=un.time)

## fN.elapsed un.elapsed
##    432.281      0.093
\end{verbatim}
}

\noindent \blu{Observe that the Woodbury trick version is more than 4000 times
faster.  This experiment was run an eight-core Intel i7 with 32 GB of RAM with
{\sf R}'s default linear algebra libraries (i.e., not the threaded MKL
library used for ATO experiment in Section \ref{sec:ato}).}

\blu{Just to check that both methods are performing the same inference, the code
chunk below prints the pairs of estimated lengthscales $\hat{\theta}$ to the
screen.}

{\singlespacing
\begin{verbatim}
R> rbind(fN=fN$theta, un=un$theta)

##        [,1]     [,2]
## fN 1.099367 1.789508
## un 1.099308 1.789700
\end{verbatim}
}

\noindent \blu{Above we specified the the {\tt X} input as a list with components {\tt X0}, {\tt Z0}
and {\tt mult} to be explicit about the mapping between unique-$n$ and
full-$N$ representations.  However, this is not the default (or intended)
mechanism for providing the design, which is to simply provide a matrix {\tt
X}.  In that case, the implementation first internally calls the
{\tt find\_reps} function in order to locate replicates in the design, and
build the appropriate list structure for further calculation.}

\section{Supporting heteroskedastic elements}
\label{sec:hetap}

\blu{Here we provide derivations supporting our claim that smoothed and optimal
solutions (for latent variables) coincide at the maximum likelihood setting.
We then provide a heteroskedastic example using the {\tt hetGP} library.
Finally, we provide an empirical demonstration illustrating the consistency of
{\tt hetGP} optimization of the latent variance variables.}

\subsection{Proof of Lemma \ref{lem:g}}
\label{ap:lemg}
\begin{proof}

Denote ($\boldsymbol{\theta}^*, \Deltan^*, \boldsymbol{\phi}^*, g^*$) as the
maximizer of Eq.~(\ref{eq:jllik}). For simplicity, we work here with variances
and not log variances. By definition, $\Lan^* = \Cg \Ug^{-1} \Deltan^*$ if and
only if $\Ug \Cg^{-1} \Lan^* = \Deltan^*$. Notice that the top part of (\ref{eq:jllik}) is invariant as long
as $\Lan = \Lan^*$. We thus concentrate on the remaining terms: $\log(\nug)$
and $\log|\Ug|$. First, the derivative of $\log|\Ug|$ with respect to $g$ is
$\tr(\An^{-1} \Ug^{-1})$, which is positive (trace of a positive definite
matrix), hence $\log|\Ug|$ increases with $g$; it also does not depend on
$\Deltan$. As for $\log(\nug)$, $n \nug = \Deltan^{*\top} \Ug^{-1} \Deltan^*
\geq \Lan^{*\top} \Cg^{-1} \Lan^*$ since
\[
\Deltan^{*\top} \Ug^{-1} \Deltan^* = \Lan^{* \top} \Cg^{-1} \Ug \Ug^{-1} \Ug \Cg^{-1} \Lan^* = \Lan^{* \top} \Cg^{-1} \Lan^* + g\Lan^{* \top} \Cg^{-1} \An^{-1} \Cg^{-1} \Lan^*
\]
and $\Cg^{-1} \An^{-1} \Cg^{-1}$ is positive definite.
Hence $g^* = 0$ and $\Lan^* = \Deltan^*$.
\end{proof}

\subsection{Illustration in {\tt hetGP}}
\label{sec:motohet}

\blu{The {\sf R} code below reproduces the key aspects our application of {\tt
hetGP} on the motorcycle data (available, e.g., in the package {\tt MASS} \cite{Venables2002}), in particular the bottom row of Figure
\ref{fig:motor2}.}

{\singlespacing
\begin{verbatim}
R> library(MASS)
R> system.time(het2 <- mleHetGP(mcycle$times, mcycle$accel, lower=15,
+    upper=50, covtype="Matern5_2"))[3]

## elapsed
##   0.641
\end{verbatim}
}

\noindent \blu{So it takes about one second to optimize over the unknown
parameters, including the latent variances, when linked to {\sf R}'s default
linear algebra libraries.  Observe that raw inputs ({\tt times}) and responses
({\tt accel}) are provided, trigging a call to the package's internal
{\tt find\_reps} function to navigate the small amount replication in the
design via the Woodbury trick.}

\blu{The plotting code below generates a pair of plots similar to those in the
final row of Figure \ref{fig:motor2}, which we do not duplicate here.  Code for
other examples, including the more expensive Monte Carlo experiments for our
ATO and SIR examples, is available upon request.}

{\singlespacing
\begin{verbatim}
R> Xgrid <- matrix(seq(0, 60, length.out = 301), ncol = 1)
R> p2 <- predict(x=Xgrid, object=het2, noise.var=TRUE)
R> ql <- qnorm(0.05, p2$mean, sqrt(p2$sd2 + p2$nugs))
R> qu <- qnorm(0.95, p2$mean, sqrt(p2$sd2 + p2$nugs))
R> par(mfrow=c(1,2))
R> plot(mcycle$times, mcycle$accel, ylim=c(-160,90), ylab="acc",
+   xlab="time")
R> lines(Xgrid, p2$mean, col=2, lwd=2)
R> lines(Xgrid, ql, col=2, lty=2); lines(Xgrid, qu, col=2, lty=2)
R> plot(Xgrid, p2$nugs, type="l", lwd=2, ylab="s2", xlab="time",
+   ylim=c(0,2e3))
R> points(het2$X0, sapply(find_reps(mcycle[,1],mcycle[,2])$Zlist, var),
+   col=3, pch=20)
\end{verbatim}
}

\subsection{Robustness in numerical optimization}
\label{ap:opt}

\blu{In Section \ref{sec:implement} we advocate initializing gradient-based
optimization of the latent noise variables using residuals obtained from
homogeneous fit, and this is the default in {\tt hetGP}.  To illustrate that
this is a good starting point, difficulties in MLE estimation for GP
hyperparameterization notwithstanding  \citep{erickson2017comparison}, we provide here
two demonstrations that this procedure is efficient. The first example reuses
the motorcycle data while the second is the 2d Branin function
\citep{dixon:szego:1978} redefined on $[0,1]^2$ with i.i.d Gaussian additive
noise with variance: $2 + 2\sin(\pi x_1)\cos(3  \pi x_2) + 5(x_1^2 +
x_2^2)))$. Unique $n=100$ design points have been sampled uniformly, along
with uniformly allocated replicates until $N=1000$.  The experiments compare
the log-likelihood and other estimates obtained via \texttt{L-BFGS-B}
optimization under our default initialization, and collection of
random ones.}

\begin{figure}[ht!]
  \centering
     Motorcycle\\
	\includegraphics[scale=0.5,trim=5 15 5 40,clip=TRUE]{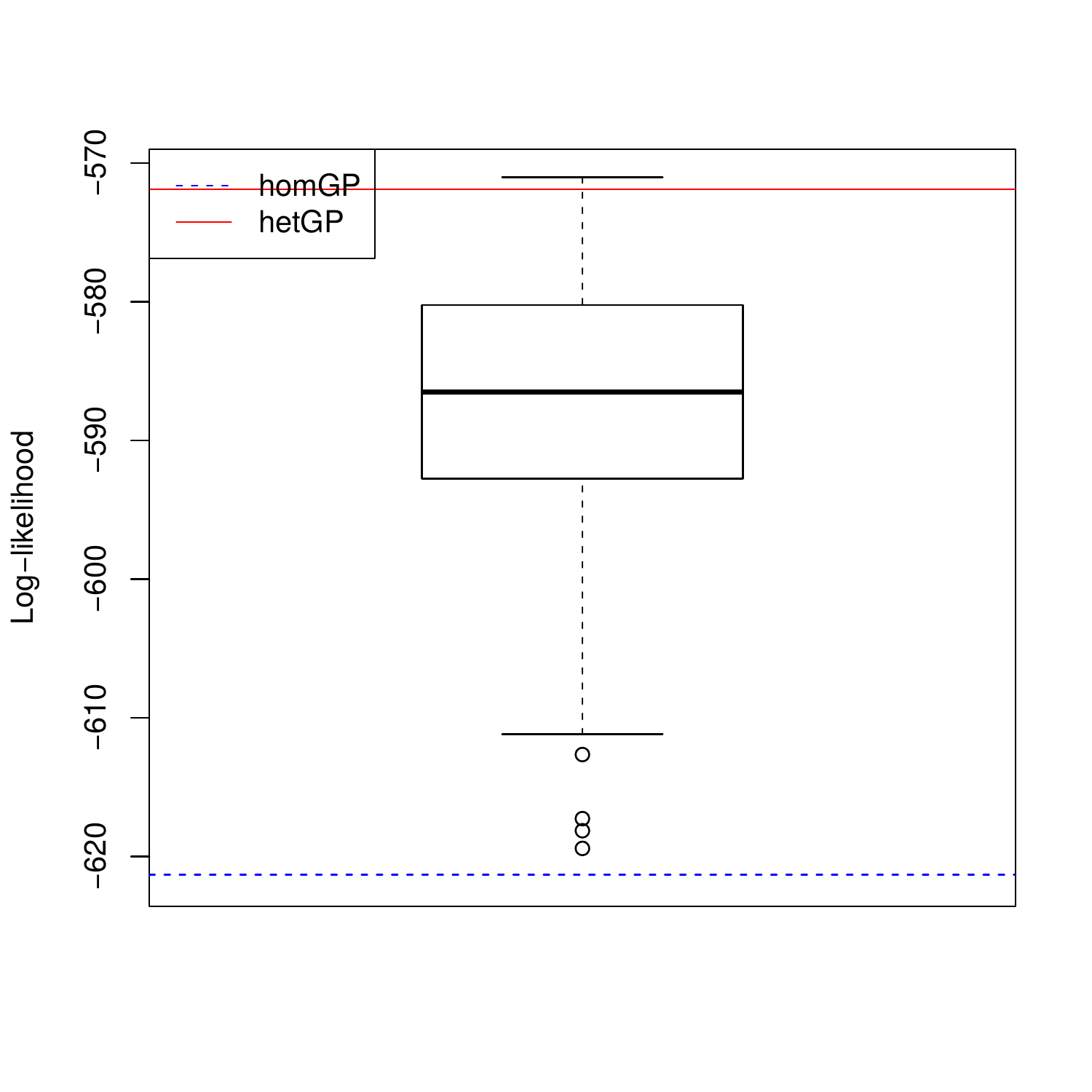}
	\includegraphics[scale=0.5,trim=5 15 5 40,clip=TRUE]{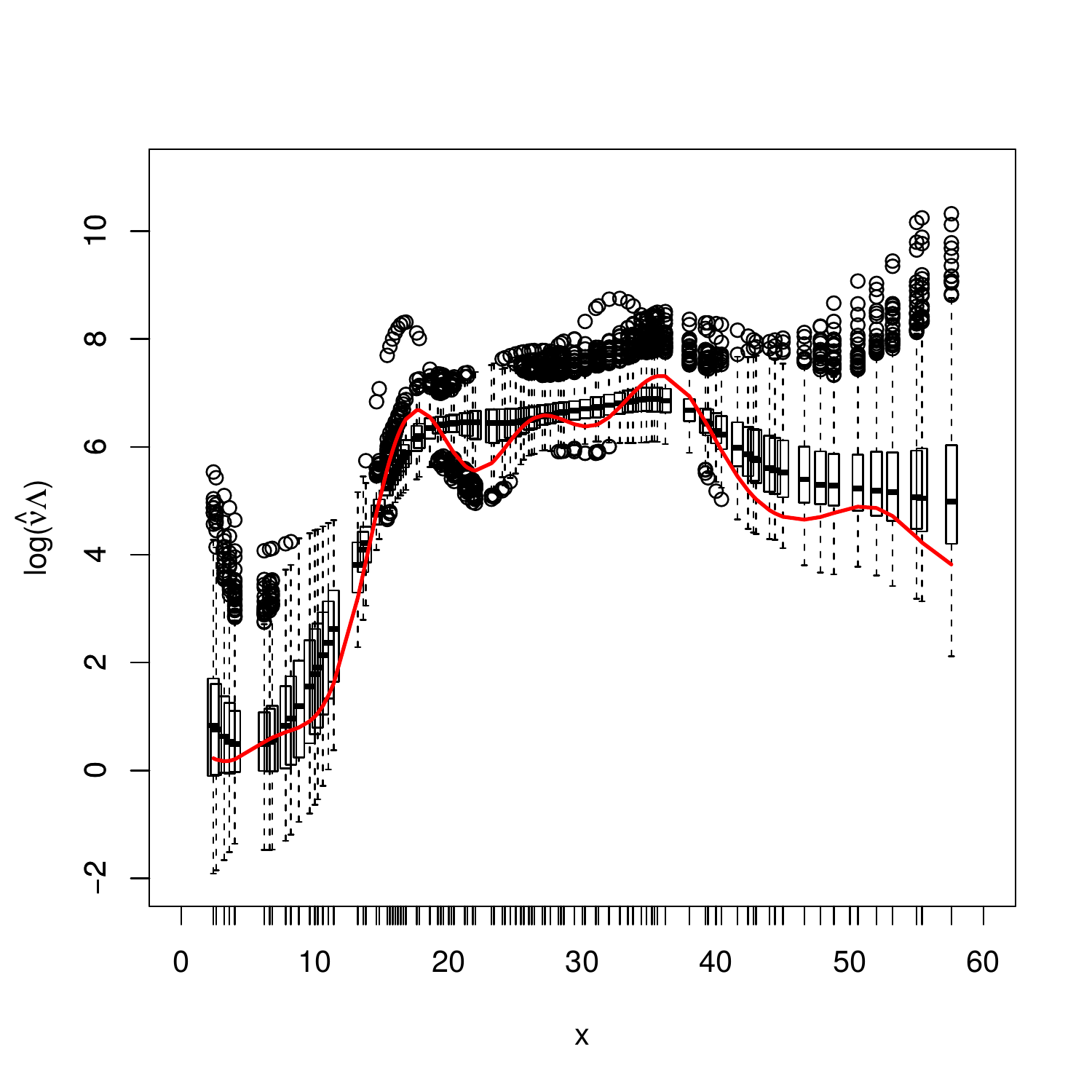}\\
	Noisy Branin \\
	\includegraphics[scale=0.5,trim=5 35 5 30,clip=TRUE]{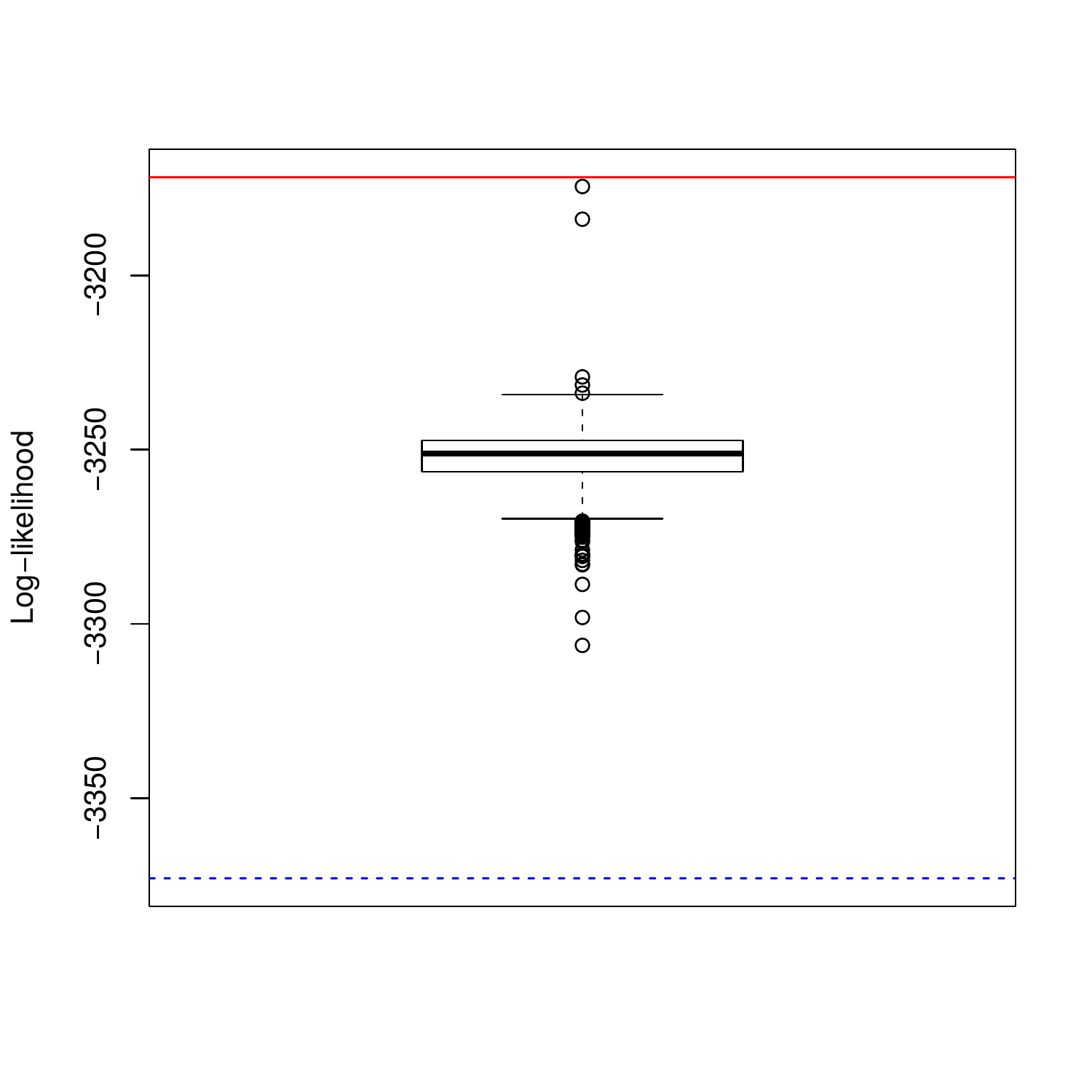}
	\includegraphics[scale=0.5,trim=5 35 5 30,clip=TRUE]{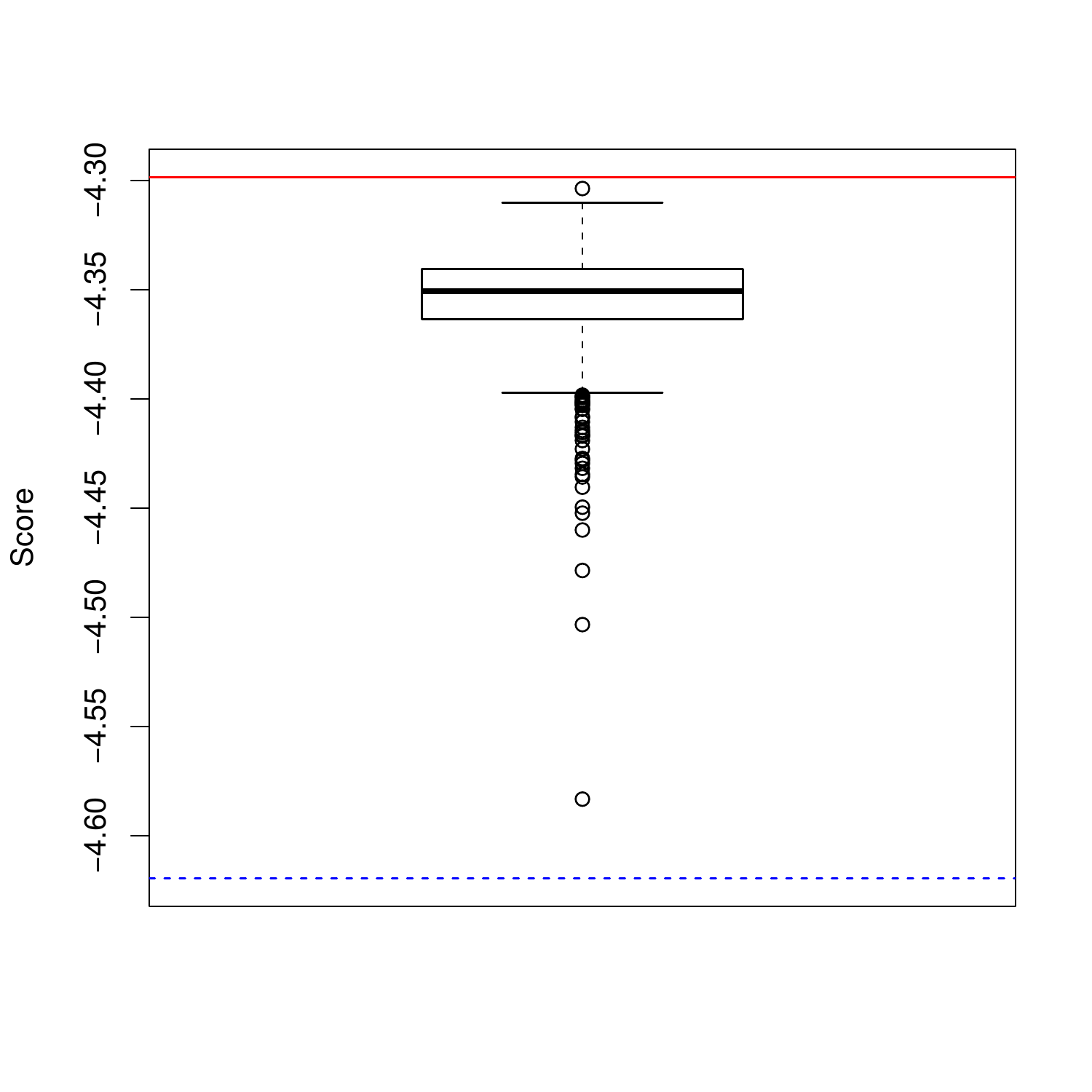}

	\vspace{-0.5cm}
	\caption{\blu{The left column compares joint MLE-values from one thousand random starting
	points to the proposed starting point, for the motorcycle data (top) and a
	2d Branin test case (bottom).  The top right panel shows boxplots for
	locally optimal $\log(\nun \Lan)$ values returned for the motorcycle data.
	The bottom right panel is a boxplot of scores for the 2d Branin example
	(based on a $51 \times 51$ grid with 10 replicates each as testing data).
	The performance of \texttt{hetGP}, i.e., under thoughtful initialization,
	is represented by the solid red in all four panels. A blue dotted line
	shows the homoskedastic fit for reference.}}
	\label{fig:testopt}
\end{figure}

\blu{The results in Figure \ref{fig:testopt} demonstrate that the proposed
optimization procedure is well calibrated. In all four panels, boxplots show
the distribution of estimates (via log likelihoods, latent variables, or
proper scores, respectively), drawing a comparison to values obtained under
our thoughtful initialization (in solid red), relative to the baseline
homoskedastic fit (blue-dashed).  More than 99\% of the time our thoughtful
initialization leads to better estimates than under random initialization.
The top-right panel, which is in log space, shows that some latent variable
estimates obtained under random initialization are quite egregiously over-estimated.
It is harder to perform such a visualization in 2d, so the bottom-right panel
shows the more aggregated score metric instead.}

\section{Non-stationarity}
\label{ap:nonstat}

\begin{figure}[ht!]
\includegraphics[scale = 0.46, trim = 0 10 30 50, clip = TRUE]{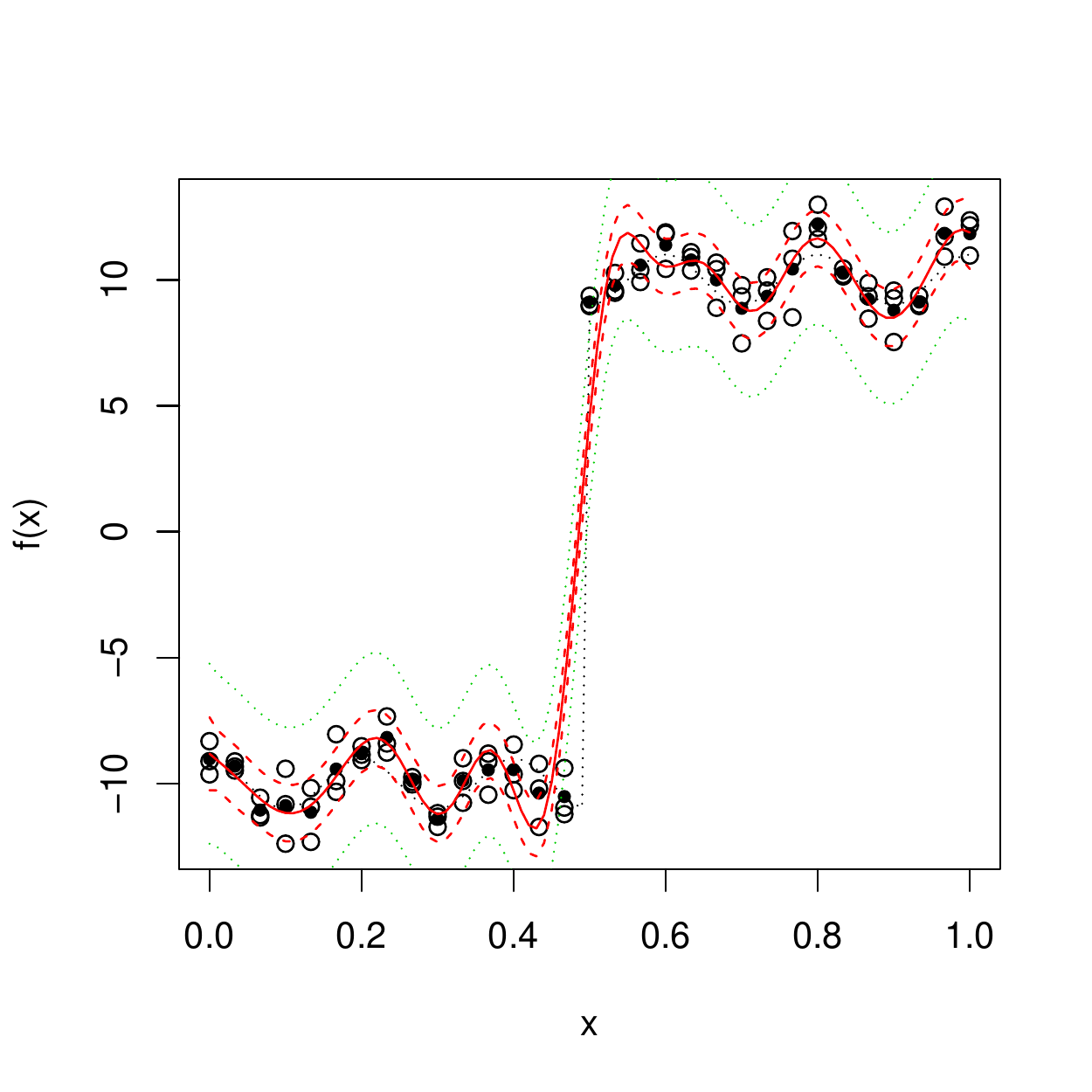}
\includegraphics[scale = 0.46, trim = 0 10 30 50, clip = TRUE]{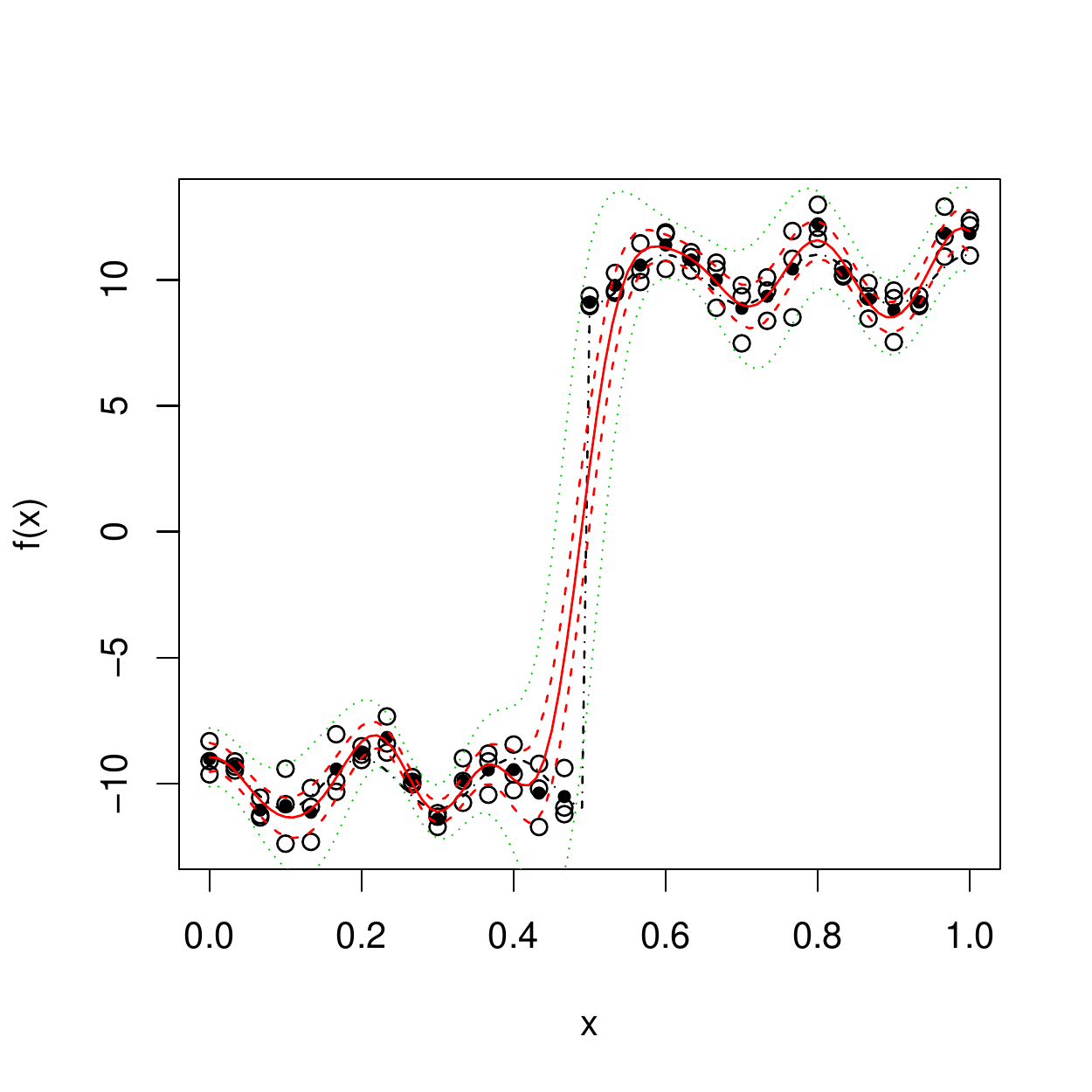}
\includegraphics[scale = 0.46, trim = 0 10 30 50, clip = TRUE]{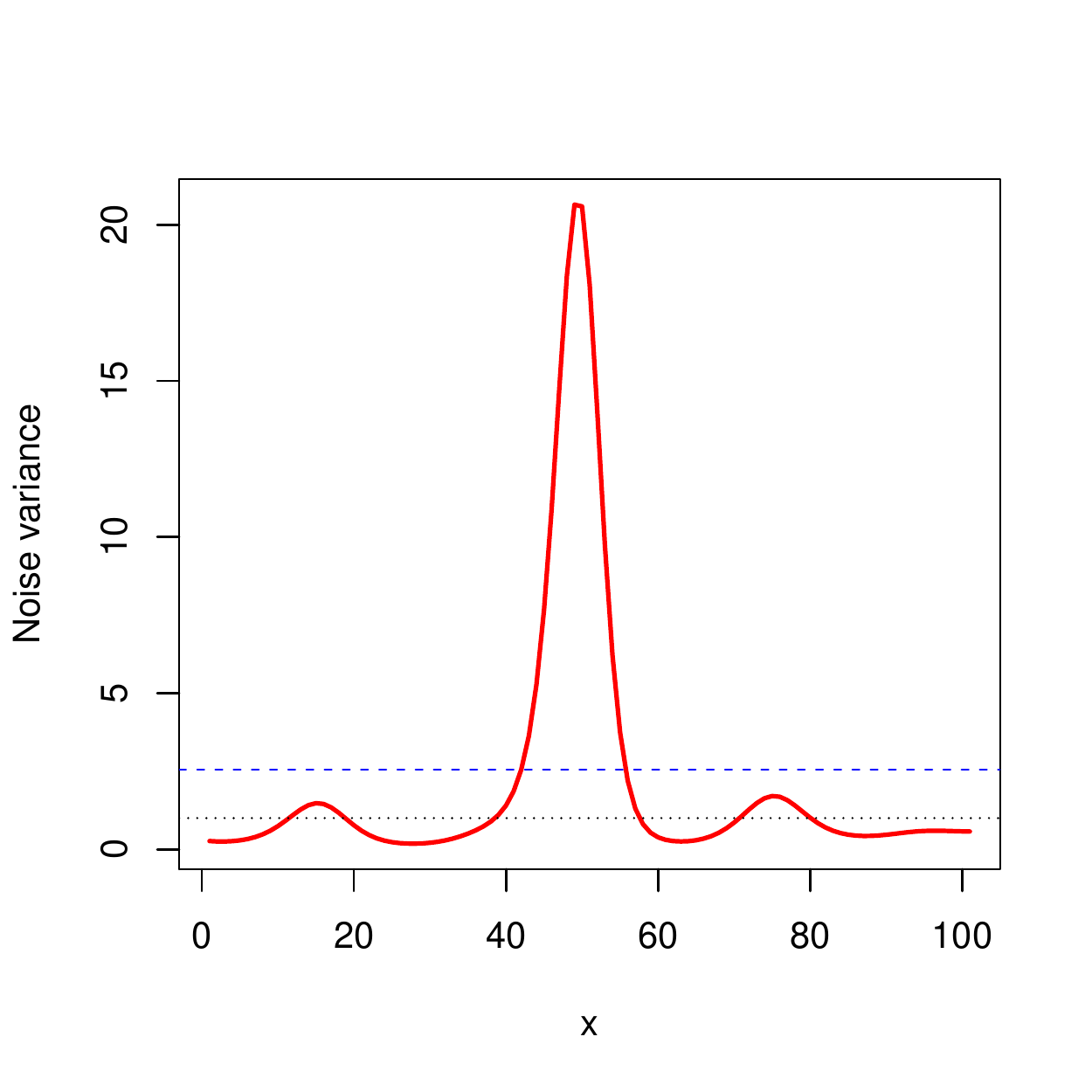}
\vspace{-0.25cm}
\caption{ \blu{Results from a homoskedastic stationary GP (left) and heteroskedastic
stationary GP (center), using the same color code as in Figure
\ref{fig:motor2} based on 31 equidistant points with 3 replicates each. Right:
estimated variance with heteroskedastic model (red solid line), with
homoskedastic model (blue dashed line) and true unit-variance (black dotted
line).}}
\label{fig:nonstat1d}
\end{figure}

\begin{figure}[ht!]
\includegraphics[scale = 0.55, trim = 10 10 10 50, clip = TRUE]{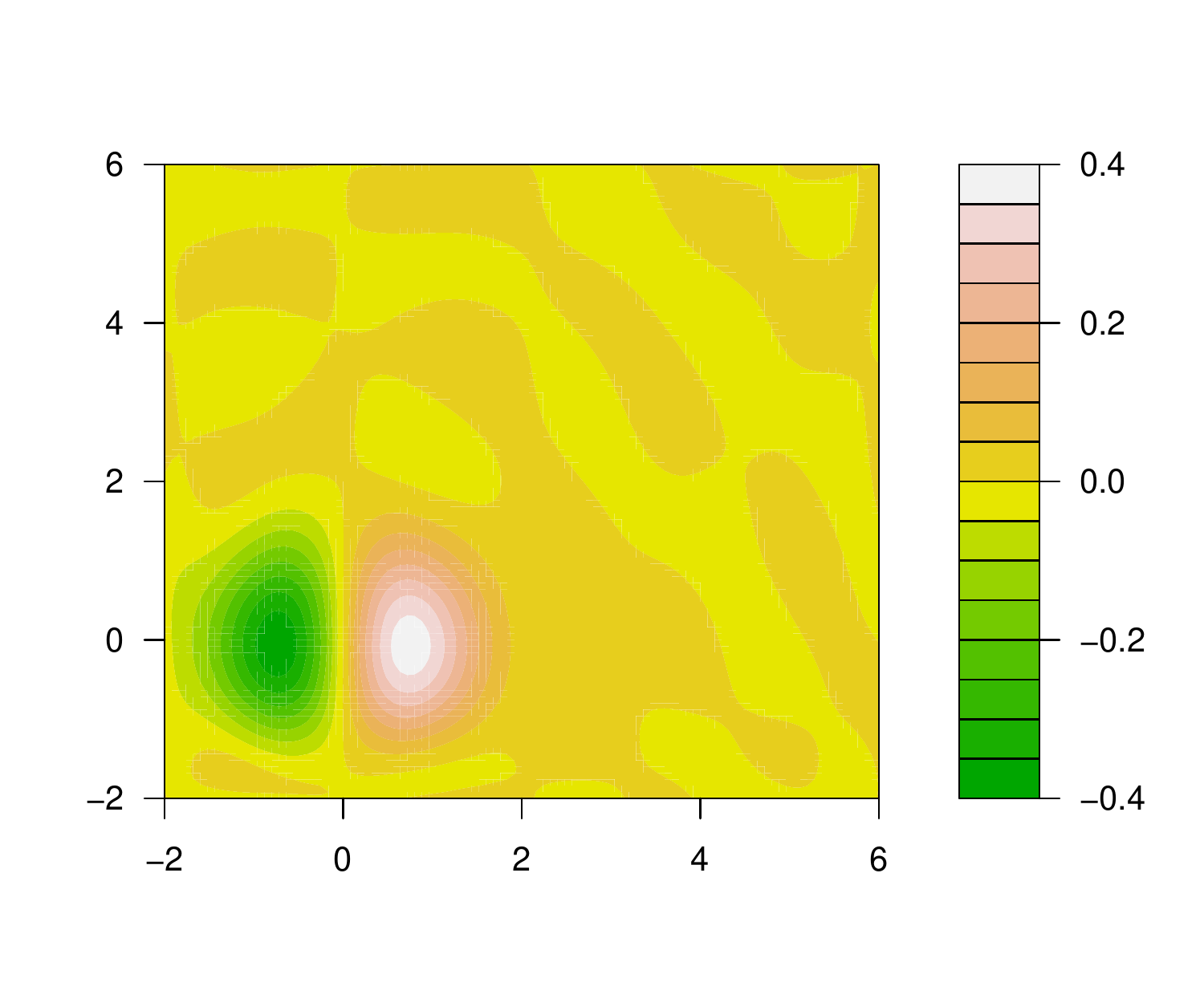}
\includegraphics[scale = 0.55, trim = 10 10 10 50, clip = TRUE]{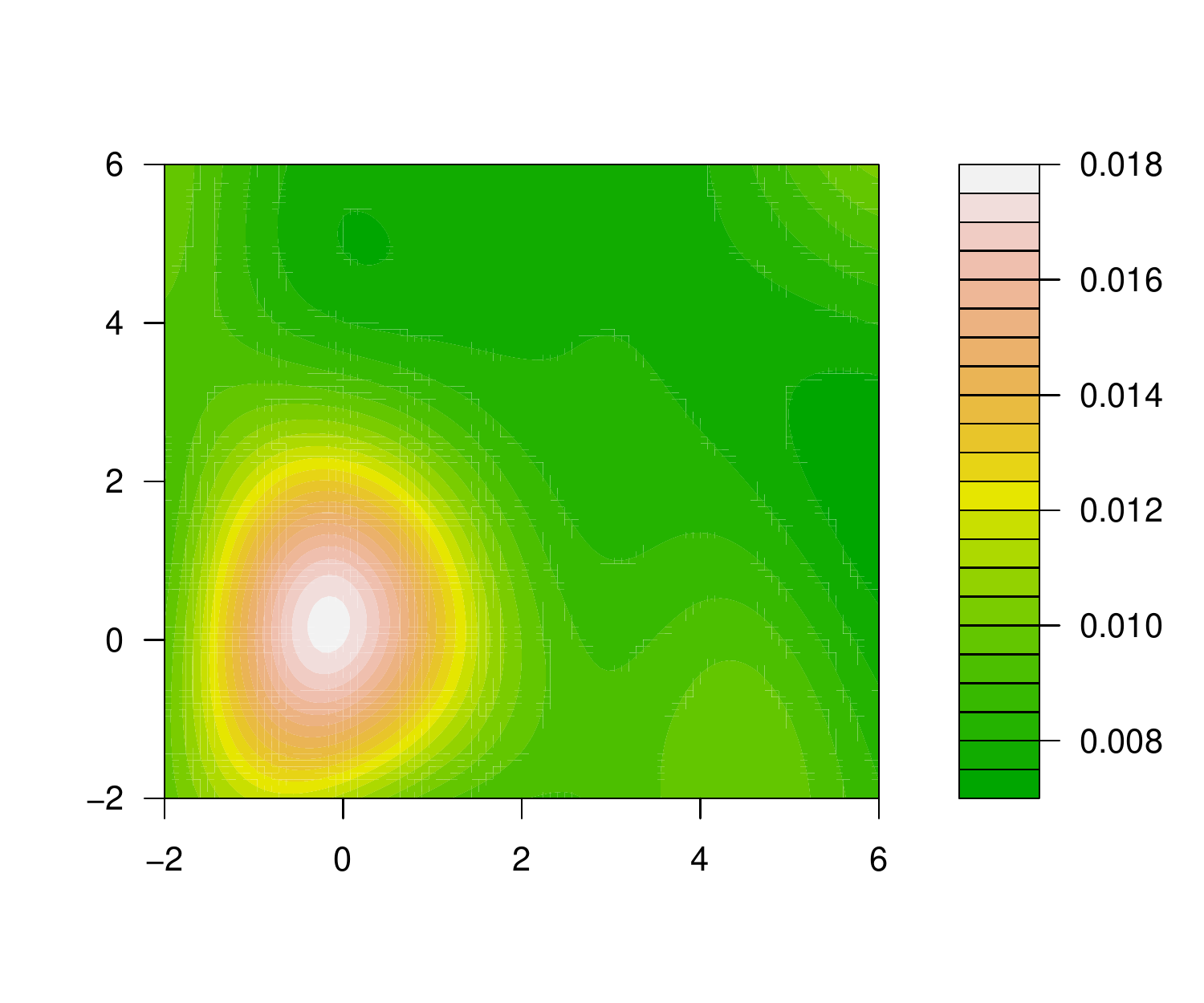}
\vspace{-0.5cm}
\caption{\blu{Estimated predictive mean (left) and standard deviation (right)
surfaces on the two-bumps function \cite{Gramacy2007}, based on a $21 \times
21$ grid with 3 replicates each. The true homogeneous standard deviation used
is $10^{-2}$.}}
\label{fig:nonstat2d}
\end{figure}

\blu{Input-dependent noise is one form of non-stationarity amongst many. Perhaps
the most commonly addressed form of non-stationary is in the mean, however the
recent literature caters to many disparate forms \citep[e.g.,][]{gramacy:lee:2008,Ba2012,Snoek2014,gramacy:apley:2015,roininen2016,marmin2017}. When there
is little data available, distinguishing between non-stationarity in the mean
and heteroskedasticity of the noise is hard, but even a mis-specified
heteroskedastic-only model may be sufficient (i.e., better than a purely
stationary one) for many tasks. We illustrate this on two examples. The first
one is an univariate sinusoidal function with a jump and the second is the
two-dimensional function of \cite{Gramacy2007}, described above in Section
\ref{sec:empwood}. The results are shown in Figures \ref{fig:nonstat1d} and
\ref{fig:nonstat2d}, respectively.}
\blu{In both cases, the estimated noise increases in areas where mean dynamics are
most rapidly changing.  This leads to improved accuracy, and can facilitate
active learning heuristics in sequential design \citep{gramacy:lee:2009}.
Focusing design effort in regions of high noise can be useful even if the true
explanation is rapidly changing signal (not high noise).  Nevertheless, when
the design increases in the ``interesting region'', either through further
exploration or a larger number of replicates, this behavior disappears and a
homogeneous model is returned. Investigations on hybrids of noise and mean
heterogeneity represents a promising avenue for future research.}

\bibliography{hetGP}

\begin{thebibliography}{41}
\newcommand{\enquote}[1]{``#1''}
\expandafter\ifx\csname natexlab\endcsname\relax\def\natexlab#1{#1}\fi

\bibitem[\protect\citename{Ankenman et~al., }2010]{ankennman:nelson:staum:2010}
Ankenman, B.~E., Nelson, B.~L., and Staum, J. (2010).
\newblock \enquote{Stochastic kriging for simulation metamodeling.}
\newblock {\em Operations Research\/}, 58, 371--382.

\bibitem[\protect\citename{Ba et~al., }2012]{Ba2012}
Ba, S., Joseph, V.~R., et~al. (2012).
\newblock \enquote{Composite Gaussian process models for emulating expensive
  functions.}
\newblock {\em The Annals of Applied Statistics\/}, 6, 4, 1838--1860.

\bibitem[\protect\citename{Banerjee et~al., }2008]{Banerjee2008}
Banerjee, S., Gelfand, A.~E., Finley, A.~O., and Sang, H. (2008).
\newblock \enquote{Gaussian predictive process models for large spatial data
  sets.}
\newblock {\em Journal of the Royal Statistical Society: Series B (Statistical
  Methodology)\/}, 70, 4, 825--848.

\bibitem[\protect\citename{Binois and Gramacy, }2017]{Binois2017}
Binois, M. and Gramacy, R.~B. (2017).
\newblock {\em {\sf hetGP}: Heteroskedastic Gaussian Process Modeling and
  Design under Replication\/}.
\newblock R package version 1.0.0.

\bibitem[\protect\citename{Boukouvalas and Cornford,
  }2009]{boukouvalas:cornford:2009}
Boukouvalas, A. and Cornford, D. (2009).
\newblock \enquote{Learning heteroscedastic {G}aussian processes for complex
  datasets.}
\newblock Tech. rep., Aston University, Neural Computing Research Group.

\bibitem[\protect\citename{{Core Team}, }2014]{R}
{Core Team}, {\sf R}. (2014).
\newblock {\em {\sf R}: A Language And Environment For Statistical
  Computing\/}.
\newblock {\sf R} Foundation for Statistical Computing, Vienna, Austria.

\bibitem[\protect\citename{Dixon and Szego, }1978]{dixon:szego:1978}
Dixon, L. and Szego, G. (1978).
\newblock \enquote{The global optimization problem: an introduction.}
\newblock {\em Towards global optimization\/}, 2, 1--15.

\bibitem[\protect\citename{Eidsvik et~al., }2014]{eidsvik2014estimation}
Eidsvik, J., Shaby, B.~A., Reich, B.~J., Wheeler, M., and Niemi, J. (2014).
\newblock \enquote{Estimation and prediction in spatial models with block
  composite likelihoods.}
\newblock {\em Journal of Computational and Graphical Statistics\/}, 23,
  295--315.

\bibitem[\protect\citename{Erickson et~al., }2017]{erickson2017comparison}
Erickson, C.~B., Ankenman, B.~E., and Sanchez, S.~M. (2017).
\newblock \enquote{Comparison of Gaussian process modeling software.}
\newblock {\em European Journal of Operational Research\/}.

\bibitem[\protect\citename{Gneiting and Raftery, }2007]{gneiting:raftery:2007}
Gneiting, T. and Raftery, A.~E. (2007).
\newblock \enquote{Strictly proper scoring rules, prediction, and estimation.}
\newblock {\em Journal of the American Statistical Association\/}, 102, 477,
  359--378.

\bibitem[\protect\citename{Goldberg et~al.,
  }1998]{goldberg:williams:bishop:1998}
Goldberg, P.~W., Williams, C.~K., and Bishop, C.~M. (1998).
\newblock \enquote{Regression with input-dependent noise: A {G}aussian process
  treatment.}
\newblock In {\em Advances in Neural Information Processing Systems\/},
  vol.~10,  493--499. Cambridge, MA: MIT press.

\bibitem[\protect\citename{Gramacy and Lee, }2009]{gramacy:lee:2009}
Gramacy, R. and Lee, H. (2009).
\newblock \enquote{Adaptive Design and Analysis of Supercomputer Experiment.}
\newblock {\em Technometrics\/}, 51, 2, 130--145.

\bibitem[\protect\citename{Gramacy, }2007]{Gramacy2007}
Gramacy, R.~B. (2007).
\newblock \enquote{tgp: an R package for Bayesian nonstationary, semiparametric
  nonlinear regression and design by treed Gaussian process models.}
\newblock {\em Journal of Statistical Software\/}, 19, 9, 6.

\bibitem[\protect\citename{{Gramacy} and {Apley}, }2015]{gramacy:apley:2015}
{Gramacy}, R.~B. and {Apley}, D.~W. (2015).
\newblock \enquote{Local Gaussian process approximation for large computer
  experiments.}
\newblock {\em Journal of Computational and Graphical Statistics\/}, 24, 2,
  561--578.

\bibitem[\protect\citename{Gramacy and Lee, }2008]{gramacy:lee:2008}
Gramacy, R.~B. and Lee, H. K.~H. (2008).
\newblock \enquote{Bayesian Treed Gaussian Process Models With an Application
  to Computer Modeling.}
\newblock {\em Journal of the American Statistical Association\/}, 103, 483,
  1119--1130.

\bibitem[\protect\citename{Haaland and Qian, }2011]{haaland:qian:2012}
Haaland, B. and Qian, P. (2011).
\newblock \enquote{Accurate emulators for large-scale computer experiments.}
\newblock {\em Annals of Statistics\/}, 39, 6, 2974--3002.

\bibitem[\protect\citename{Harville, }1997]{harville:1997}
Harville, D. (1997).
\newblock {\em Matrix algebra from a statistician's perspective\/}.
\newblock Springer-Verlag.

\bibitem[\protect\citename{Hong and Nelson, }2006]{hong:nelson:2006}
Hong, L. and Nelson, B. (2006).
\newblock \enquote{Discrete optimization via simulation using {COMPASS}.}
\newblock {\em Operations Research\/}, 54, 1, 115--129.

\bibitem[\protect\citename{Hu and Ludkovski, }2017]{hu2015sequential}
Hu, R. and Ludkovski, M. (2017).
\newblock \enquote{Sequential design for ranking response surfaces.}
\newblock {\em SIAM/ASA Journal on Uncertainty Quantification\/}, 5, 1,
  212--239.

\bibitem[\protect\citename{Kami{\'n}ski, }2015]{kaminski:2015}
Kami{\'n}ski, B. (2015).
\newblock \enquote{A method for the updating of stochastic Kriging metamodels.}
\newblock {\em European Journal of Operational Research\/}, 247, 3, 859--866.

\bibitem[\protect\citename{Kaufman et~al., }2012]{kaufman:etal:2012}
Kaufman, C., Bingham, D., Habib, S., Heitmann, K., and Frieman, J. (2012).
\newblock \enquote{Efficient emulators of computer experiments using compactly
  supported correlation functions, with an application to cosmology.}
\newblock {\em Annals of Applied Statistics\/}, 5, 4, 2470--2492.

\bibitem[\protect\citename{Kersting et~al., }2007]{kersting:etal:2007}
Kersting, K., Plagemann, C., Pfaff, P., and Burgard, W. (2007).
\newblock \enquote{Most likely heteroscedastic {G}aussian process regression.}
\newblock In {\em Proceedings of the International Conference on Machine
  Learning\/},  393--400. New York, NY: ACM.

\bibitem[\protect\citename{Lazaro-Gredilla and Titsias,
  }2011]{lazaro-gredilla:tsitas:2011}
Lazaro-Gredilla, M. and Titsias, M. (2011).
\newblock \enquote{Variational heteroscedastic {G}aussian process regression.}
\newblock In {\em Proceedings of the International Conference on Machine
  Learning\/},  841--848. New York, NY: ACM.

\bibitem[\protect\citename{Lee and Owen, }2015]{lee:owen:2015}
Lee, M.~R. and Owen, A.~B. (2015).
\newblock \enquote{Single Nugget Kriging.}
\newblock Tech. rep., Stanford University.
\newblock ArXiv:1507.05128.

\bibitem[\protect\citename{Ludkovski and Niemi, }2010]{ludkovski:niemi:2010}
Ludkovski, M. and Niemi, J. (2010).
\newblock \enquote{Optimal dynamic policies for influenza management.}
\newblock {\em Statistical Communications in Infectious Diseases\/}, 2, 1,
  article 5.

\bibitem[\protect\citename{Marmin et~al., }2017]{marmin2017}
Marmin, S., Ginsbourger, D., Baccou, J., and Liandrat, J. (2017).
\newblock \enquote{Warped Gaussian processes and derivative-based sequential
  design for functions with heterogeneous variations.}

\bibitem[\protect\citename{Merl et~al., }2009]{merl:etal:2009}
Merl, D., Johnson, L.~R., Gramacy, R.~B., and Mangel, M. (2009).
\newblock \enquote{A statistical framework for the adaptive management of
  epidemiological interventions.}
\newblock {\em PloS One\/}, 4, 6, e5807.

\bibitem[\protect\citename{Ng and Yin, }2012]{ng:yin:2012}
Ng, S.~H. and Yin, J. (2012).
\newblock \enquote{Bayesian kriging analysis and design for stochastic
  systems.}
\newblock {\em ACM Transations on Modeling and Computer Simulation (TOMACS)\/},
  22, 3, article no.~17.

\bibitem[\protect\citename{Nychka et~al., }2015]{nychka:etal:2015}
Nychka, D., Bandyopadhyay, S., Hammerling, D., Lindgren, F., and Sain, S.
  (2015).
\newblock \enquote{A Multi-resolution {G}aussian process model for the analysis
  of large spatial data sets.}
\newblock {\em Journal of Computational and Graphical Statistics\/}, 24, 2,
  579--599.

\bibitem[\protect\citename{Opsomer et~al., }1999]{opsomer:etal:1999}
Opsomer, J., Ruppert, D., Wand, W., Holst, U., and Hossler, O. (1999).
\newblock \enquote{Kriging with nonparameteric variance function estimation.}
\newblock {\em Biometrics\/}, 55, 704--710.

\bibitem[\protect\citename{Picheny and Ginsbourger,
  }2013]{picheny:ginsbourger:2013}
Picheny, V. and Ginsbourger, D. (2013).
\newblock \enquote{A nonstationary space-time {G}aussian process model for
  partially converged simulations.}
\newblock {\em SIAM/ASA Journal on Uncertainty Quantification\/}, 1, 57--78.

\bibitem[\protect\citename{Plumlee, }2014]{plumlee:2014}
Plumlee, M. (2014).
\newblock \enquote{Efficient inference for random fields using sparse grid
  designs.}
\newblock {\em Journal of the American Statistical Association\/}, 109, 508,
  1581--1591.

\bibitem[\protect\citename{Quadrianto et~al., }2009]{quadrianto:etal:2009}
Quadrianto, N., Kersting, K., Reid, M., Caetano, T., and Buntine, W. (2009).
\newblock \enquote{Kernel conditional quantile estimation via reduction
  revisited.}
\newblock In {\em Proceedings of the 9th IEEE International Conference on Data
  Mining\/},  938--943.

\bibitem[\protect\citename{Roininen et~al., }2016]{roininen2016}
Roininen, L., Girolami, M., Lasanen, S., and Markkanen, M. (2016).
\newblock \enquote{Hyperpriors for Mat{\'e}rn fields with applications in
  Bayesian inversion.}
\newblock {\em arXiv preprint arXiv:1612.02989\/}.

\bibitem[\protect\citename{Roustant et~al.,
  }2012]{roustant:ginsbourger:deville:2012}
Roustant, O., Ginsbourger, D., and Deville, Y. (2012).
\newblock \enquote{{DiceKriging}, {DiceOptim}: Two {R} Packages for the
  analysis of computer experiments by kriging-based metamodeling and
  optimization.}
\newblock {\em Journal of Statistical Software\/}, 51, 1, 1--55.

\bibitem[\protect\citename{Snelson and Ghahramani,
  }2005]{snelson:ghahramani:2005}
Snelson, E. and Ghahramani, Z. (2005).
\newblock \enquote{Sparse Gaussian processes using pseudo-inputs.}
\newblock In {\em Advances in Neural Information Processing Systems\/},
  1257--1264.

\bibitem[\protect\citename{Snelson and Gharamani,
  }2006]{snelson:ghahramani:2006}
Snelson, E. and Gharamani, Z. (2006).
\newblock \enquote{Variable noise and dimensionality reduction in {G}aussian
  processes.}
\newblock In {\em Proceedings of the International Conference in Artificial
  Intelligence\/}.
\newblock Also see ar{X}iv:1506.04000.

\bibitem[\protect\citename{Snoek et~al., }2014]{Snoek2014}
Snoek, J., Swersky, K., Zemel, R.~S., and Adams, R.~P. (2014).
\newblock \enquote{Input Warping for Bayesian Optimization of Non-stationary
  Functions.}
\newblock In {\em International Conference on Machine Learning\/},  1674--1682.

\bibitem[\protect\citename{Venables and Ripley, }2002]{Venables2002}
Venables, W.~N. and Ripley, B.~D. (2002).
\newblock {\em Modern Applied Statistics with S\/}.
\newblock 4th ed. New York: Springer.
\newblock ISBN 0-387-95457-0.

\bibitem[\protect\citename{Wang and Chen, }2016]{Wang2016}
Wang, W. and Chen, X. (2016).
\newblock \enquote{The effects of estimation of heteroscedasticity on
  stochastic kriging.}
\newblock In {\em Proceedings of the 2016 Winter Simulation Conference\/},
  326--337. IEEE Press.

\bibitem[\protect\citename{Xie et~al., }2012]{xie:frazier:chick:2012}
Xie, J., Frazier, P., and Chick, S. (2012).
\newblock \enquote{Assemble to Order Simulator.}

\end{thebibliography}
\bibliographystyle{jasa}

\end{document}